\documentclass[aps,pra,twocolumn,superscriptaddress,10pt,showpacs]{revtex4}
\usepackage[dvips]{graphics}
\usepackage{graphicx}
\usepackage{amsfonts}
\usepackage{amssymb}
\usepackage{amsmath}
\usepackage{subfigure}

\begin{document}

\title{Localized structures in Kagom{\'e} lattices}
\author{K.\ J.\ H.\ Law}
\affiliation{Department of Mathematics and Statistics,
University of Massachusetts,
Amherst MA 01003-4515, USA}
\author{Avadh Saxena}
\affiliation{Theoretical Division and Center for Nonlinear Studies,
Los Alamos National Laboratory, Los Alamos, NM 87545, USA
}
\author{P.\ G.\ Kevrekidis}
\affiliation{Department of Mathematics and Statistics,
University of Massachusetts,
Amherst MA 01003-4515, USA}
\author{A.\ R.\ Bishop}
\affiliation{Theoretical Division and Center for Nonlinear Studies,
Los Alamos National Laboratory, Los Alamos, NM 87545, USA
}

\begin{abstract}
We investigate the existence and stability of gap vortices and
multi-pole gap solitons in a Kagom{\'e} lattice with a defocusing
nonlinearity both in a discrete
case and in a continuum one with periodic external modulation.
In particular, predictions are made based on expansion around
a simple and analytically tractable anti-continuum (zero coupling) limit.
These predictions are then confirmed for a continuum model of an
optically-induced Kagom{\'e} lattice in a photorefractive crystal
obtained by a continuous transformation of a honeycomb lattice.

%In particular, we consider it as an appropriate transformation
%of the honeycomb lattice by tilting the beams and modulating
%one.

\end{abstract}

\maketitle

\section{Introduction}
\label{intro}

Recent years have seen a great deal of interest in Hamiltonian lattice
and quasi-discrete systems due to their relevance as
models of experiments coming from various branches of physics.
An early example is that of the nonlinear optics of fabricated AlGaAs
waveguide arrays \cite{7}.  The interplay of discreteness and nonlinearity
there led to the emergence of numerous phenomena that
have gathered considerable attention subsequently, such as Peierls-Nabarro
potential barriers, diffraction and diffraction management \cite{7a} and
gap solitons \cite{7b}, to name just a few.  See, for example,
the reviews \cite{review_opt,general_review} and references therein.

More recently, there has been growing interest within nonlinear
optics in the area of optically-induced photonic lattices in photorefractive
crystals such as strontium barium niobate (Sr$_x$Ba$_{1-x}$NbO$_3$, commonly abbreviated SBN).  The original
theroetical  proposal of \cite{efrem} was followed quickly by experimental
realizations \cite{moti1,moti2}, and the
foundation was thus set for the observation of a diverse array of
novel nonlinear phenomena in this setting.  Among others, these phenomena
include dipole \cite{dip}, multipole \cite{quad}, necklace \cite{neck},
gap \cite{7b}
and rotary \cite{rings} solitons as well as discrete \cite{vortex1,vortex2}
and gap \cite{motihigher}
vortices, higher order Bloch modes \cite{neshev2},
Zener tunneling \cite{zener}, as well as localized modes
in honeycomb \cite{honey}, hexagonal \cite{rosberg2}
and quasi-crystalline \cite{motinature1} lattices,
and Anderson localization \cite{motinature2}
(see, e.g., the reviews \cite{moti3,zc} for additional examples).

A considerable effort along these lines has been dedicated to
the recently emerging area of {\it non-square} lattices
\cite{ablo,seg,ol2007a,ol2007b,honey,gaid,
rosberg2,szameit,kouk,tja2,bam,ourhon,ourhexhon,moti07,IK}.
 Furthermore, the majority of these studies have dealt with the case
of a focusing non-linearity rather than a defocusing one.  Coherent
structures in the latter case have received relatively limited
attention until very recently, for example in the study of fundamental
and higher order gap solitons \cite{moti1}.
More complicated gap structures, such as multipoles and complex valued
vortices are only starting to be explored in square lattices
\cite{ourol,ourox}.  A theoretical
framework has been developed in parallel 
to this work, stemming from one-dimensional
and square lattices \cite{ourpre,pgk_dnls}.
However, the predictions of the latter
can also be translated to contours (or paths) in
non-square geometries \cite{ourhon,ourhexhon}, based on 
arguments of dimensionality reduction along the contour.

In this work we will focus on the so-called
{\it Kagom{\'e}} lattice, which is encountered often in nature and has
a very rich structure.  In the solid-state community and other areas
of physics and science these lattices and many others have been
%thoroughly
explored for decades \cite{something}, but are becoming more prevalent
recently \cite{kag1,kag2}.
Furthermore, low temperature properties of atomic quantum (ultracold Bose
and Fermi) gases have been studied in the trimerized Kagom{\'e} lattice
\cite{something}.
%They also describe laser arrangements to create such a lattice.

Motivated by recent advances in optically-induced lattices in
SBN,
%Sr$_x$Ba$_{1-x}$NbO$_3$,
we will explore a Non-linear Schr{\"o}dinger (NLS)
model in both its discrete (DNLS) manifestation as a set of difference equations
adhering to the symmetry of the lattice and modeling coupled oscillators, and in
the analogous continuum setting using a partial differential equation with an external
potential having the appropriate symmetry.  In particular, since the continuum
model is motivated by experiments with SBN,
%Sr$_x$Ba$_{1-x}$NbO$_3$,
the nonlinearity
will be {\it saturable} \cite{moti3,zc}.  
We will investigate prototypical contours (or paths)
of localized structures in this lattice, consisting of six sites as well as
four sites, and being both real, and complex valued with continuous phase
(modulo $2\pi$).

Our main findings in what follows are that

\begin{itemize}
\item Certain structures are stable, such as the in-phase gap hexapole and
single-charge six-site gap vortex on the honeycomb cell, and the
in-phase/out-of-phase quadrupole
on the ``hourglass cell" (see Fig. \ref{F1}). Other configurations are unstable.
\item In the continuum model, continuations of solutions
in the first band-gap pass through
the second band as quasi-localized %``embedded'' 
structures and
then become fully extended in the second band-gap. However, discontinuous
extensions, i.e. new continuations of the localized structures,
are found to exist
in the second band-gap simultaneously with the extended states.
\item The result of the evolution of the dynamical instability
in these lattices is more complex than in the square lattice case,
and may involve not
%just
only degeneration
to single-site solitons but
%in the discrete case
possibly to multi-site solitary wave structures,
and, in the discrete case,
often the formation of robust breathing states, consisting of
multiple sites (possibly even as many as in the original configuration).
In fact, we have found some clear breather formations recurring in multiple simulations:
\begin{itemize}
\item Two nearest-neighbor or opposite sites in-phase with each other and
with oscillating amplitudes of comparable magnitude.
\item Two next-nearest-neighbor sites out-of-phase with each other and
with oscillating amplitudes of comparable magnitude.
\item Two nearest-neighbor sites having different amplitudes
and oscillating between the same
phases and opposite phases depending on whether the
amplitudes are further from or closer to each other, respectively.
\end{itemize}
And, in the continuum version, either all or most of the initially populated
wells remain populated for long propagation distances, with the instability manifesting
itself only as phase reshaping.
\end{itemize}

The presentation will be structured as follows.  In section \ref{problem}
we provide the setup of the problem, the background and the theory.  Then in
section \ref{numerics} we will systematically explore the relevant numerical
results.  Finally, in section \ref{conclusion} we will summarize our findings
and present our conclusions.

\section{Setup}
\label{problem}

The description of the setup of the problem, including the background and
theory is organized into three sections: first, the preliminary
material \ref{prelims}; second the existence considerations
\ref{existence}; and third the stability considerations \ref{stability}.

\subsection{Preliminaries}
\label{prelims}

\begin{figure}[tbh]
\begin{center}
\includegraphics[width=0.4\textwidth]{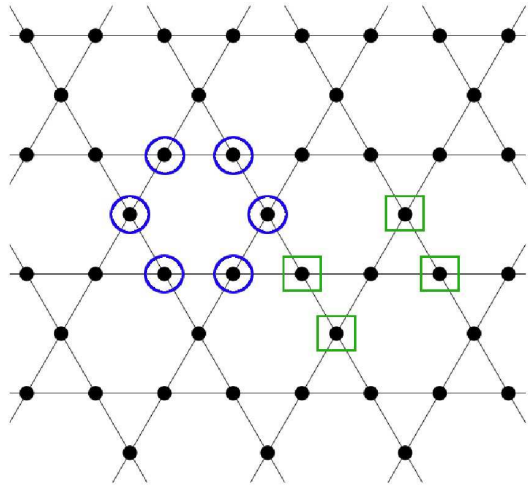}
\end{center}
\caption{(Color online) The discrete Kagom{\'e} lattice structure is presented
above.
The six-site contour is given by the blue circles, while the
four-site ``hourglass'' contour is given by the green squares.}
\label{F1}
\end{figure}

We introduce the following complex-valued non-linear
evolution equation
%for the amplitude of the electric field $U$
%\cite{ol2007a,ol2007b,yang04_3,yang04_4},
%in the following form:

\begin{equation}
-iU_z=[\mathcal{L}+\mathcal{N}(\textbf{x},|U|^2)]U,\label{eq1}
\end{equation}

\noindent where $U$ is a function of $z \in \mathbb{R}_+$ and $\textbf{x} \in
\mathbb{Z} \times \mathbb{Z}$ in the discrete version or else
$\textbf{x} \in \mathbb{R} \times \mathbb{R}$ in the continuum version.
First, we will consider the discrete version with

\begin{equation}
\mathcal{L}=\epsilon \left ( \sum_{j \in \{\pm J\}} \textbf{e}_j - 4 \right ),
\end{equation}

\noindent where $\textbf{e}_J$ is a translation by one site in the positive
direction along $J$ and $J$ is one of a site-dependent subset of two of the
principal lattice vectors $\textbf{a}_d=(1,\sqrt{3})/2$,
$\textbf{b}_d=(-1,\sqrt{3})/2$, or $\textbf{c}_d=(1,0)$
%along which the two indices in
%$\mathbb{Z} \times \mathbb{Z}$ define the
of the
discrete Kagom{\'e} lattice
presented in Fig. \ref{F1} and
$\epsilon$ is the coupling between sites.
The non-linear term is taken
to be a cubic Kerr non-linearity as follows

\begin{equation}
\mathcal{N}(\textbf{x},|U|^2)=-|U|^2.
\label{eq1_d}
\end{equation}

The simulations for the static results in the
discrete model were performed in the domain
$D_h \backslash K$, where
$D_h=[1, \dots, 33] \times [1, \dots, 33]$
is the discrete lattice domain corresponding
to a triangular lattice and
$K=\{(2m+1,2n+1) | (m,n)\in [0,16]^2 \}$.
For dynamical evolution, the solutions were
buffered with $40$ (or more) nodes on all
sides to prevent radiation scattering from
the boundaries.

The continuum version consists of defining
$\mathcal{L}=D \nabla^2$, where $\nabla^2$
is the two-dimensional Laplacian on
$\mathbb{R} \times \mathbb{R}$ and

\begin{equation}
\mathcal{N}(\textbf{x},|U|^2)=\frac{E_0}{1+I(\textbf{x})+|U|^2},
\label{eq1_c}
\end{equation}

\noindent with

\begin{equation}
I(\textbf{x})=
I_0\left|f_1(\textbf{x}) e^{ik\textbf{b}_1 \cdot \textbf{x}}+
{e}^{ik\textbf{b}_2 \cdot \textbf{x}}+
{e}^{ik\textbf{b}_3 \cdot \textbf{x}}\right|^2
\label{eq2}
\end{equation}
the optical lattice intensity function formed by three (p=0) or
four laser beams with
$f_1(\textbf{x})=e^{ik p x/(1+4 p/3)} \cos [p k x/(1+4 p/3)]$,
$\textbf{b}_1=(1/(1+4 p/3),0),$
$\textbf{b}_2=(-\frac{1}{2(1+4 p/3)},-\frac{\sqrt{3}}{2}),$
$\textbf{b}_3=(-\frac{1}{2(1+4 p/3)},\frac{\sqrt{3}}{2}).$
As $p \rightarrow 3/2$ the lattice transforms from the
well-known honeycomb interference pattern into
the richer Kagom{\'e} lattice. The latter lattice features both the
hexagons from the honeycomb lattice and the equilateral
triangles from the triangular one, and each node has
four neighbors similar to the square lattice.

\begin{figure}[tbh]
\begin{center}
\includegraphics[width=0.4\textwidth]{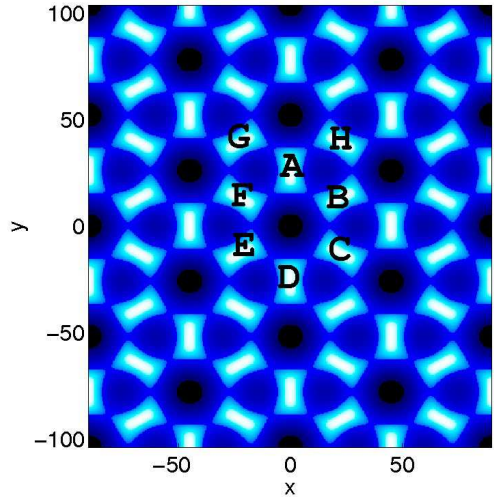}
\end{center}
\caption{(Color online) A spatial (x-y) contour plot of the
effective potential created by the ordinary polarization standing
wave [lattice beam in Eq.\ (\ref{eq2})].
Points $A,\, B,\, C,\, D,\, E,\, F, \, G$ and $H$
define the relevant potential minima for the various
configurations we will consider.
The contour $\{A,B,C,D,E,F\}$ is the honeycomb
cell, which can be considered to tile part of the lattice.
The set of sites $\{B,F,G,H\}$
comprise the ``hourglass" cell contour
we will consider.  Together with $A$, these
sites comprise another cell which tiles the
remaining part of the lattice.}
\label{F2}
\end{figure}

Here $I_0$ is the lattice peak intensity, $z$ is the propagation
constant and $\textbf{x}=(x, y)$ are transverse distances
(normalized to $z_s=1$ mm and $x_s=y_s=1 \mu$m),
$E_0$ is proportional to the applied DC field voltage,
$D=z_s \lambda/(4\pi n_e x_s y_s)$ is the diffraction coefficient,
$\lambda$ is the wavelength of the laser in a vacuum,
$d=4\pi/k$ is the periodicity in the x-direction
($d/\sqrt{3}$ is the periodicity in the y-direction)
and $n_e$ is the refractive index along the extraordinary axis.
We choose the lattice intensity $I_0 = 1$, and
$(d,E_0,\lambda,n_e)$=(90,8,532 $\textrm{nm}$,2.35), consistent
with a typical experimentally accessible situation
\cite{ol2007a,ourol}.
A plot of the potential intensity field created by the
optical lattice is shown in Fig. \ref{F2} to
illustrate the locations where our
localized configurations will live.
The non-dimensional value $D=18.01$,
and we note that this dispersion coefficient is equivalent to
rescaling space by a factor $\sqrt{D}$ as, e.g. in \cite{our_sq}.

The numerical simulations are performed in a rectangular
$120 \times 120$ grid corresponding to the
domain size $ 4d \times 8 d/ \sqrt{3}$
(i.e. four periods of the lattice in each direction),
using a rectangular spatial mesh with
$\Delta x = 1.5$ and $\Delta y \approx 1.732$.
Regarding the typical dynamics of a solution when it is unstable,
we simulate the z-dependent evolution
using a Runge-Kutta fourth-order scheme with a step
size $\Delta z=0.01$.

\subsection{Existence considerations}
\label{existence}

\begin{figure}[tbp!]
\begin{center}
\includegraphics[width=.55\textwidth]{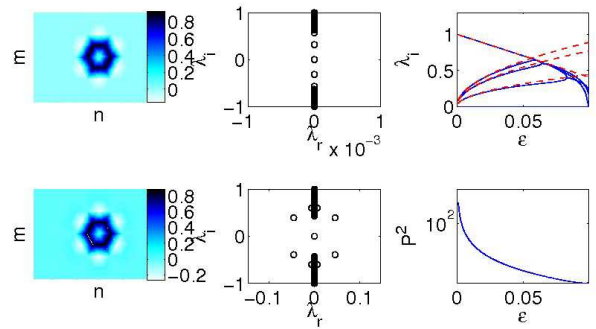}
\end{center}
\caption{(Color online) The discrete in-phase hexapole solutions
are presented.  In the first two columns the profiles
(left)
and linearization spectra are given before (top, $\epsilon=0.061$)
and after ($\epsilon=0.085$) the
first Hamiltonian Hopf (HH) bifurcation.
The top right panel depicts the theoretical
predictions of the linearization eigenvalues bifurcating from the
AC limit (dashed)
as well as the actual numerically computed ones (solid).  The 
bottom right panel is $P^2$ (see Eq. (\ref{power})), 
shown on a ${\rm log}$ scale,
where we can observe the decrease in the effective power, 
as the coupling strength increases.}
\label{ip6d}
\end{figure}

\begin{figure}[tbp!]
\begin{center}
\includegraphics[width=.55\textwidth]{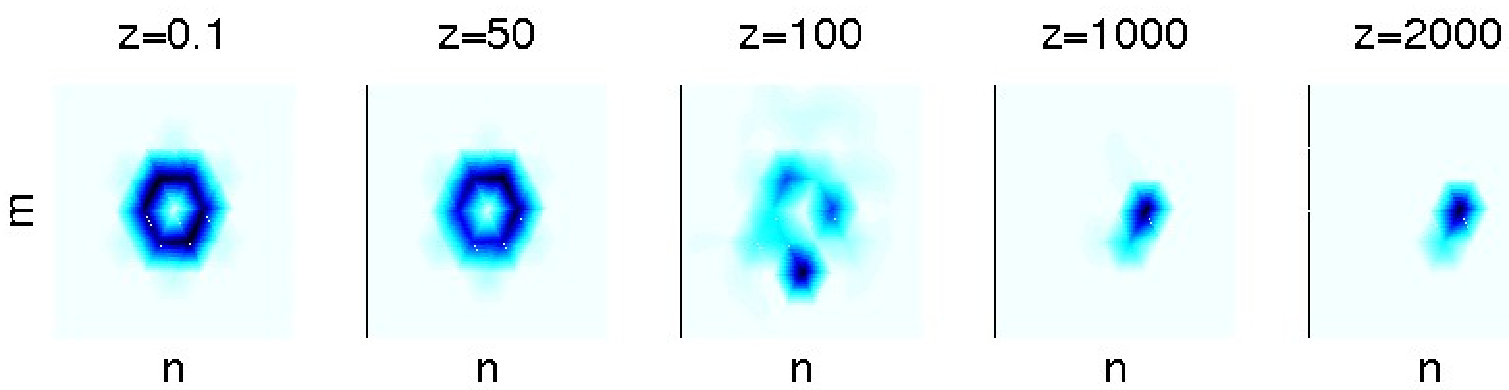}\\
\includegraphics[width=.55\textwidth]{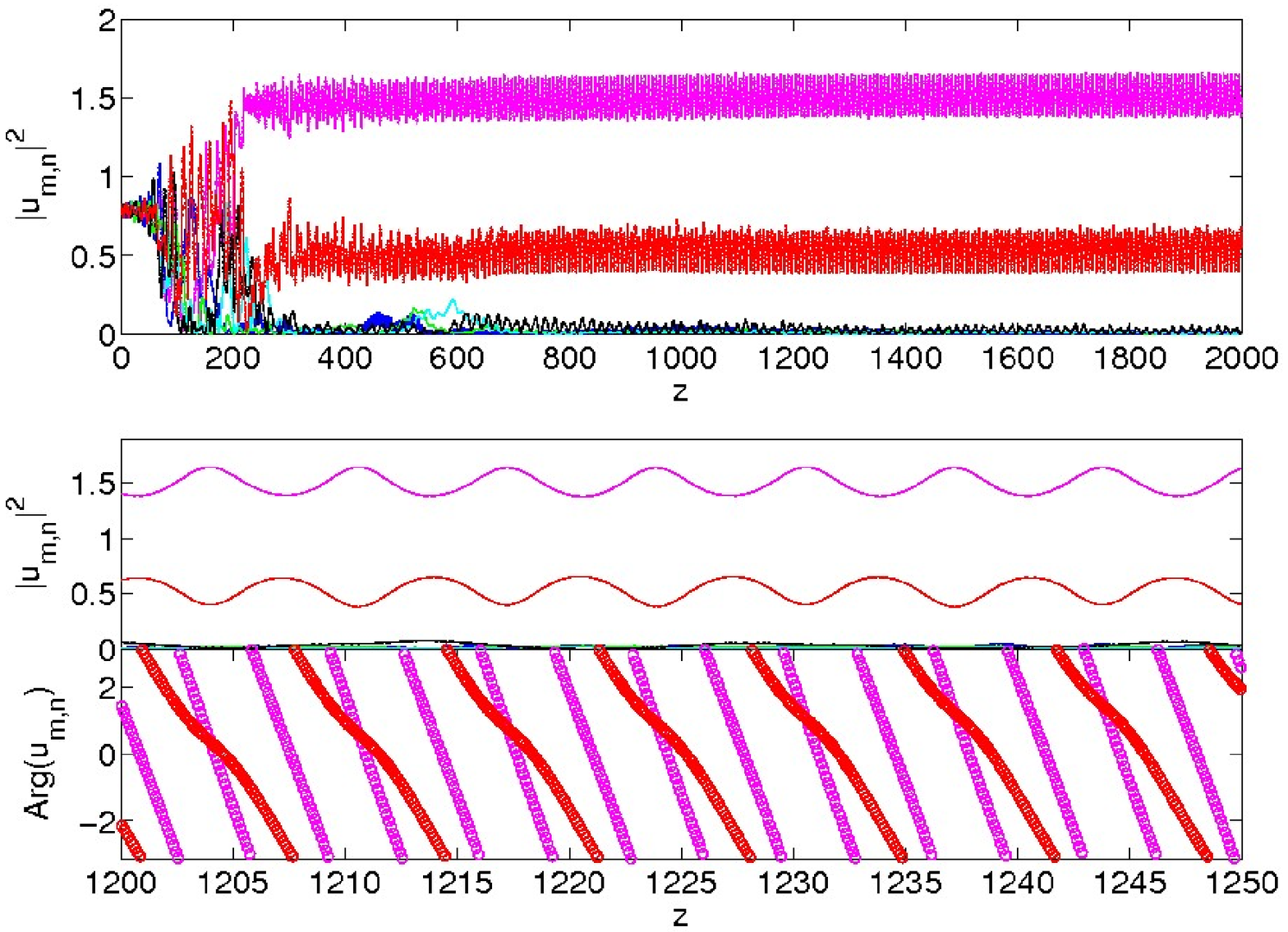}
\end{center}
\caption{(Color online) The dynamics of the solution given
in the bottom row of Fig. \ref{ip6d} is presented.
The top row shows snapshots of the modulus for various $z$,
while the next row shows the individual amplitudes at the relevant
sites. The structure survives for a while but ultimately disintegrates,
due to the instability, into two populated nearest-neighboring
sites whose amplitudes breathe closer and
further from one another while the phases oscillate between opposite
and same, respectively (see the bottom two rows).}
\label{ip6di}
\end{figure}

\begin{figure}[tbp!]
\begin{center}
\includegraphics[width=.5\textwidth]{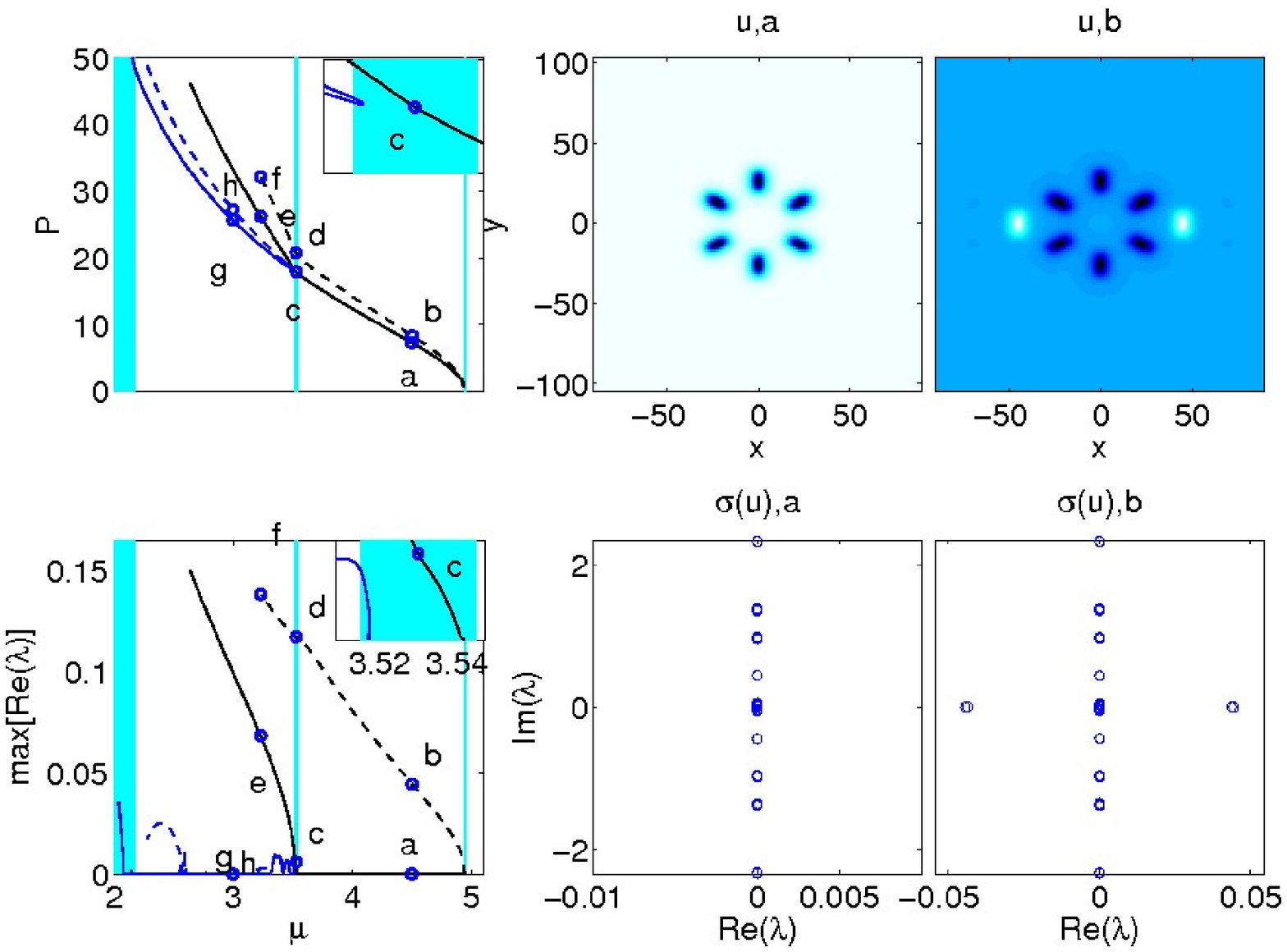}\\
\includegraphics[width=.5\textwidth]{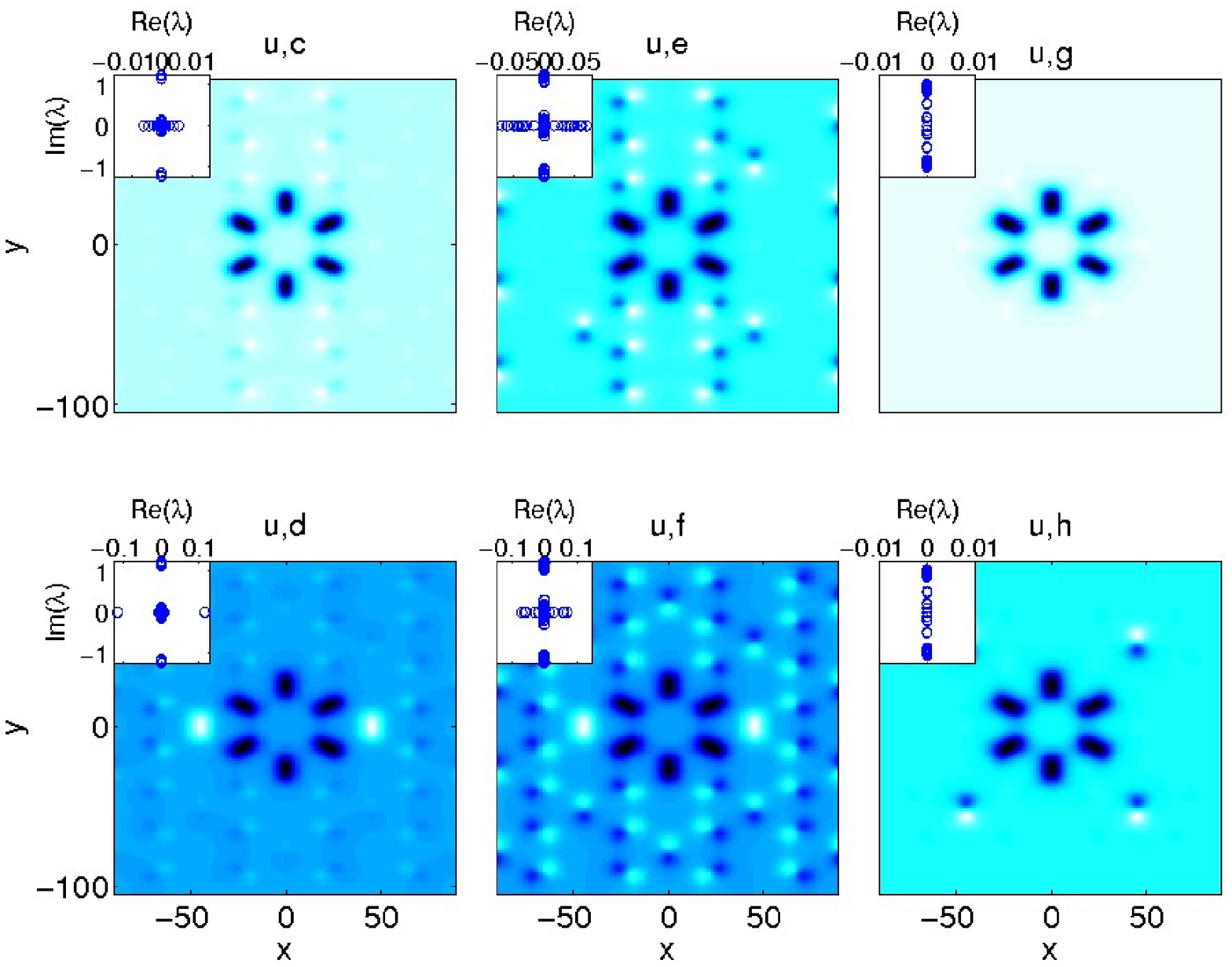}
\end{center}
\caption{(Color online) The continuum in-phase hexapole is
presented in these panels.  The top two left panels show the
power, $P$ (top) and instability 
growth rate (bottom) as given by the maximum real
part of the linearization spectrum.  The first band is given to the
right, beyond which is the semi-infinite gap (it is displayed wider
than it actually is for visibility, because its actual width is
narrower than a pixel at this scale), the second band is in the middle,
and the third band is at the far left. The blue branches in the 
second gap are actually discontinuous extensions of the localized 
modes from the first gap, which collide in a saddle-node bifurcation
and disappear as can be observed in the inset panels 
in the upper right corners (this is consistent throughout 
the following images, but the closeups will not be shown).
Solutions marked on these
plots with the letters a,b,c,d,e,f,g and h are presented in the remaining
panels.  The second and third columns of the top set display the principal
two solutions a and b, respectively, with full panels of their linearization
spectra below them.  The bottom six panels have miniature sub-panels with the
corresponding spectra embedded in them.  There are stable first (a), and
also second (g and h) gap soliton structures.  The solitons 
(c,d), with energy in the second band, are unstable.}
\label{ip6c}
\end{figure}

\begin{figure}[tbp!]
\begin{center}
\includegraphics[width=.55\textwidth]{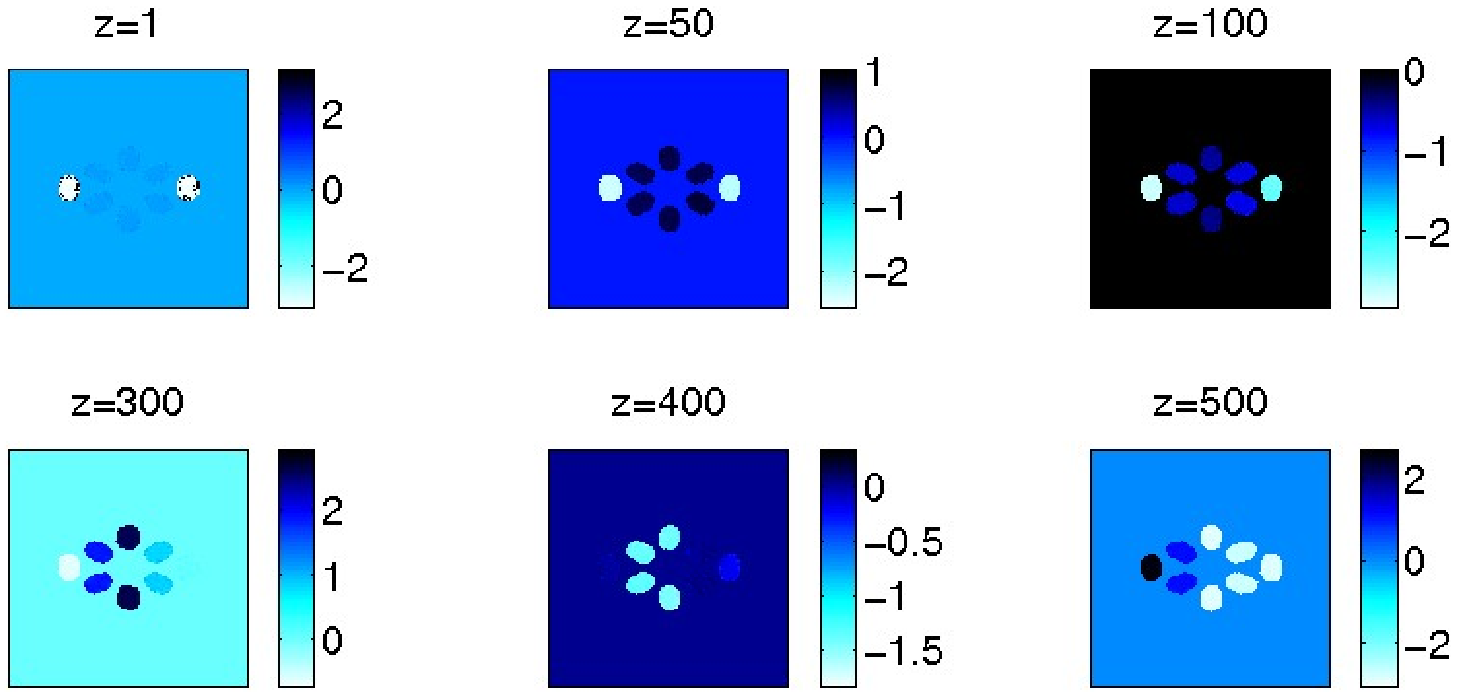}
\end{center}
\caption{(Color online) The evolution of the unstable solution
given in Fig. \ref{ip6c} (b) is shown.  The phase is shown
as ${\rm arg}(u) \chi \{(x,y) | |u|^2>0.5 {\rm max}_{(x,y)}[|u|^2] \}$ at
various times because the original configuration is preserved for a
very long propagation distance ($\chi$ is the indicator function which
%suppresses
annihilates the field outside of the set on which it is defined).
The relative phases of the configuration
break up after $z=100$.}
\label{ip6dy}
\end{figure}

Assuming a stationary state $u$ exists, and
letting the propagation constant $-\mu$
represent the (nonlinear)
real eigenvalue of the operator of the right-hand-side
of Eq. (\ref{eq1}),
the corresponding eigenvector $u$ is a fixed point of

\begin{equation}
[\mu+\mathcal{L}+\mathcal{N}(\textbf{x},|u|^2)]u=0.
\label{sta_eq}
\end{equation}

In the discrete case, we
%fixed $\mu=1$ and
performed a continuation
in $\epsilon$.  In particular, if one indexes the sites by
$(m,n)$, then solutions in the limit
$\epsilon \rightarrow 0$ can easily be found of the general form
$u_{m,n}=\sqrt{\mu}\exp{\{- i (\mu t-\theta_{m,n}) \}}$
for any arbitrary $\theta_{m,n} \in [0,2\pi)$ \cite{pgk_dnls}.
We can linearize Eq. (\ref{sta_eq})
around the solution for $\epsilon=0$ denoted by $u_0$,
accounting for complex valued perturbations by considering
the conjugate of Eq. (\ref{sta_eq}) as well, which
has the solution $u_0^*$.
The {\it Jacobian} of (\ref{sta_eq}) for $\varepsilon=0$,
or equivalently, in the absence of $\mathcal{L}$,
is 

\begin{equation}
\mathcal{J}(u)=
[\mu+ \partial(\mathcal{N}u,[\mathcal{N}u]^*)/\partial(u,u^*)].
\label{discrete_lin1}.
\end{equation}

We may also take the coupling $\varepsilon$, when sufficiently small, as
the small parameter in the expansion with $[u,u^*]^T=[u_0,u_0^*]^T+\varepsilon u_1$.
If we denote by $\mathcal{L}_{\varepsilon}$ the operator
$\mathcal{L}$ for coupling $\varepsilon$, then the first order correction
in $\varepsilon$ to Eq. (\ref{sta_eq}) is

\begin{center}
\begin{math}\label{discrete_lin2}
\mathcal{J}(u_0) u_1 +\frac{1}{\varepsilon}
\bordermatrix{& \cr & \mathcal{L}_{\varepsilon} & 0\cr
 & 0 & \mathcal{L}_{\varepsilon} \cr}
 \bordermatrix{& \cr & u_0 \cr & u_0^* \cr} =0.
\end{math}
\end{center}

Projecting this map onto the kernel of $\mathcal{J}(u_0)$
eliminates the first term and we are left with the condition

\begin{eqnarray*}
%\begin{center}
%\begin{math}
\left \langle \bordermatrix{& \cr & \mathcal{L}_{\varepsilon} & 0\cr
 & 0 & \mathcal{L}_{\varepsilon} \cr}
 \bordermatrix{& \cr & u_0 \cr & u_0^* \cr},
 {\rm ker} \{\mathcal{J}(u_0)\} \right \rangle =0,
\label{discrete_lin3}
%\end{math}
%\end{center}
\end{eqnarray*}

\noindent where we use $\langle \cdot, \cdot \rangle$ the standard
inner product on the Hilbert space $l^2$.
%\mathbb{C}^N$.
We let $\mu=1$ without loss of generality
and denote by $j$ the indices $(m,n)$
along the one-dimensional contour. The non-trivial part of the
Jacobian $\mathcal{J}(u_0)$ decouples into a direct
sum of $N$ $2 \times 2$ blocks if there are $N$ excited sites
in the contour.
For each $j$
there is a nontrivial element
$(e^{i \theta_j}, -e^{-i \theta_j})^T \in {\rm ker} \{\mathcal{J}_j(u_0)\}$.
So, the condition for existence
%, Eq. (\ref{discrete_lin3}),
of solutions to Eq. (\ref{sta_eq}) with $\varepsilon>0$
reduces to the vanishing of the vector function
${\bf g}({\boldsymbol \theta})$ of the
phase vector %${\mathbf \theta}=(\theta_1, \dots, \theta_N)$
${\boldsymbol \theta}=(\theta_1, \dots, \theta_N)$
where

\begin{equation}
g_j \equiv -\sin(\theta_j-\theta_{j-1})-\sin(\theta_j-\theta_{j+1}),
\label{persist}
\end{equation}

\noindent subject to periodic boundary conditions.
We consider primarily contours $M$
within the subcategory of
those for which
$|\theta_{j+1}-\theta_j|=\Delta \theta$
is constant for all $j \in M$,
$|\theta_1-\theta_{|M|}|=\Delta \theta$ and
$\Delta \theta |M| = 0$ ${\rm mod}$ $2\pi$.
A standard Newton fixed-point solver is used to
construct branches of solutions to Eq. (\ref{sta_eq})
in $\varepsilon$ from the AC limit.

For the continuum problem where ${\bf x} \in \mathbb{R}^2$, there
exists no such analytical solution from which to construct
a continuation.  On the other hand, it is well-known that localized
solutions exist for values of the propagation constant $\mu$
in the complement of the linear spectrum (ie. the so-called spectral gap)
defined by the following eigenvalue problem (also known as the linearization
around the zero solution),

\begin{equation}
[\mu-\mathcal{L}-\mathcal{N}(\textbf{x},0)]u=0.
\label{lin}
\end{equation}

These solutions are exponentially localized in space, so-called
gap-soliton, states of the original
nonlinear partial differential equation.
Since the parameter of interest is $\mu$, diagnostics are
plotted against $\mu$.  Continuations in this parameter
can be found with a fixed-point
solver and an initial guess of a collection of Gaussian
wave-packets in the appropriate configuration.
Using a standard eigenvalue solver package
implemented through MATLAB, we identified the first two spectral
gaps for our given parameters and grid size to be
$G_1 \approx (3.545,4.9454)$ and $G_2 \approx (2.178,3.515)$.
It is worth noting that the bands and gaps remain very close
to the same widths for much smaller discretizations
(ie. much larger grids).  For instance, with $300$ nodes in
each direction we have $\tilde{G}_2=(2.125,3.463)$, 
%and
%$|\tilde{G}_2-G_2| \sim 10^{-3}$, 
so the
change in width of the band-gaps is an order of magnitude
closer to convergence than the position (modulo translation).
%The width of the bands changes even less than this.
We use the
bands appropriate to the discretization in order to compare
them with the bifurcation structure of solutions.

The localized states $u$ of the continuum version of (\ref{sta_eq})
were obtained using the Newton-Krylov fixed point solver nsoli from
\cite{kell03}, which utilizes a GMRES iterative algorithm,
based on residual reduction in successive Krylov subspaces,
in order to minimize the memory necessary for the linear solver within
each step of the Newton algorithm.  
Some care has to be taken to handle the large size of the representation
of a 2D continuum domain.
A pseudo-arclength continuation \cite{doedel} was
used to follow each branch and locate the bifurcations which occur
at the edges of the bands.

The square root of the optical power of
the  localized waves is defined as follows:
\begin{equation}
P = \left[\int |U|^2\,dS \right]^{1/2},
\label{power}
\end{equation}

\noindent where in the continuum problem, $dS=dxdy$, 
while in the discrete problem we define the corrsponding sum 
(divided 
%for computational convenience 
by $\sqrt{\varepsilon}$).
%while in the discrete problem
%we define $dS = \delta_{m,n}/\sqrt{\varepsilon}$ with $\delta$ the Kronecker
%delta function, implying a summatio.
%\footnote{While the more natural choice for
%the weight might be $1/\varepsilon$, we use the square-root because it becomes
%very large
%%as it is
%for very small $\varepsilon$ and the discrete
%power diagrams must already be presented on a $\log$ scale.
%%We do believe
%It is important
%to represent some such measure inverse to the coupling, since the
%analogy of smaller coupling must be
%with large power solutions far from the band edge
%in the continuum model.}

\subsection{Stability considerations}
\label{stability}

\begin{figure}[tbp!]
\begin{center}
\includegraphics[width=.55\textwidth]{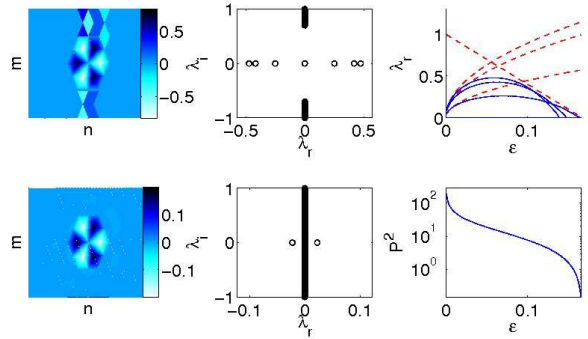}
\end{center}
\caption{(Color online) The same panels as Fig. \ref{ip6d}
except for the unstable out-of-phase hexapole.  The top
row solution is for $\varepsilon=0.05$, while the bottom one is
for $\varepsilon=0.16$.}
\label{op6d}
\end{figure}

\begin{figure}[tbp!]
\begin{center}
\includegraphics[width=.55\textwidth]{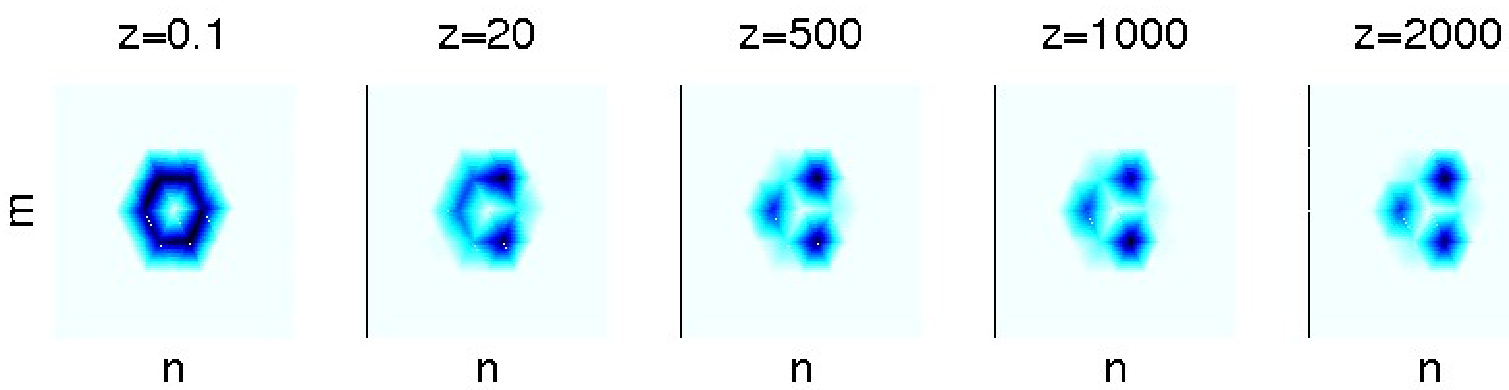}\\
\includegraphics[width=.55\textwidth]{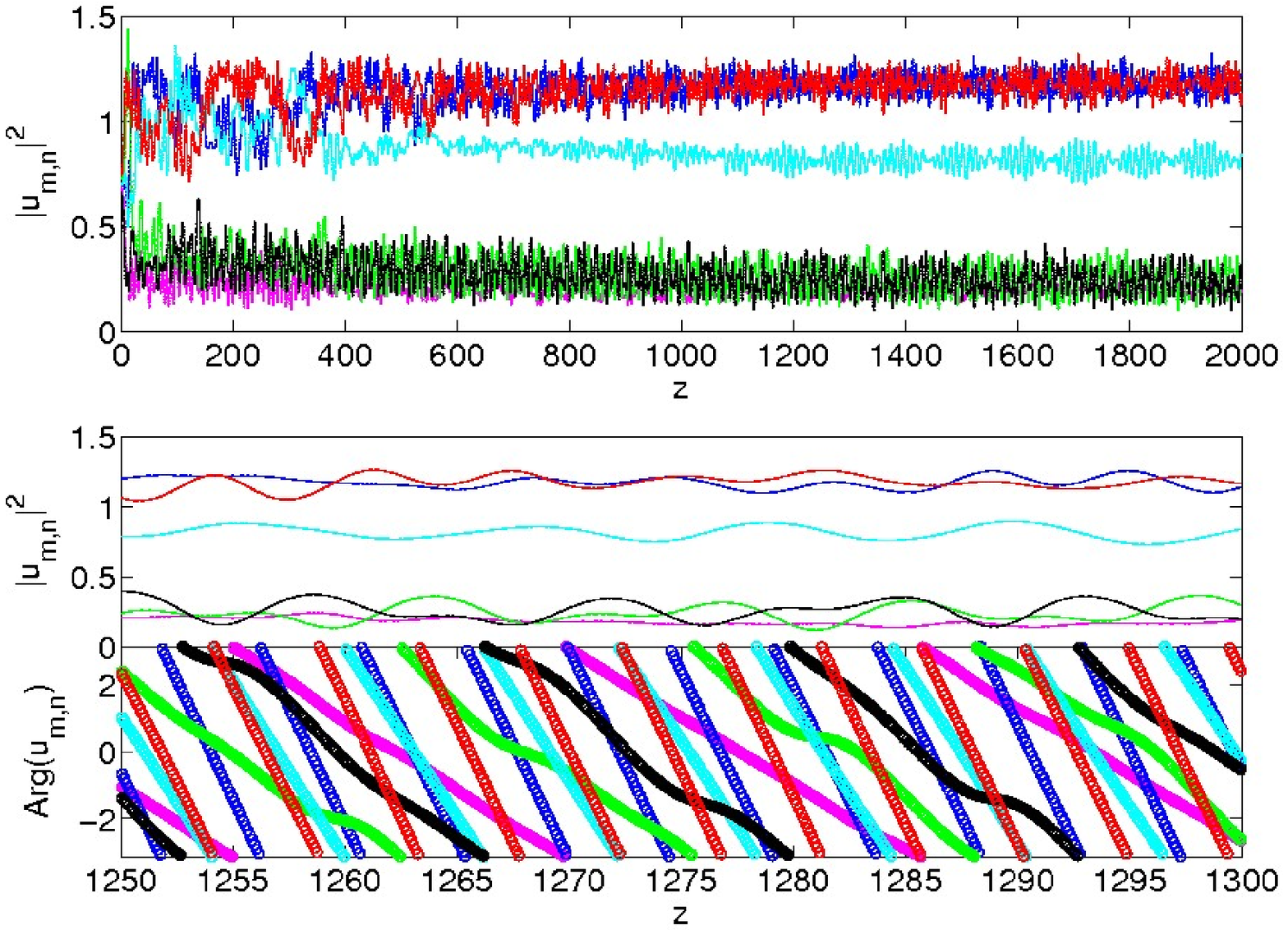}
\end{center}
\caption{(Color online)  The same as Fig. \ref{ip6di} for the
out-of-phase solution given in the top row of Fig. \ref{op6d}.
The original configuration clearly shifts very rapidly, although
the different sites remain populated for a long 
propagation distance.  The largest amplitude
pair which is separated by one node (i.e. next-nearest neighbors)
remains very close to exactly out-of-phase, while the next
smallest, which is also next-nearest neighbor to both, passes
from the phase of one to the other.  The other three sites do
not exhibit any phase correlation, although at times two have
matching phase and are out-of-phase with the other.}
\label{op6di}
\end{figure}

\begin{figure}[tbp!]
\begin{center}
\includegraphics[width=.5\textwidth]{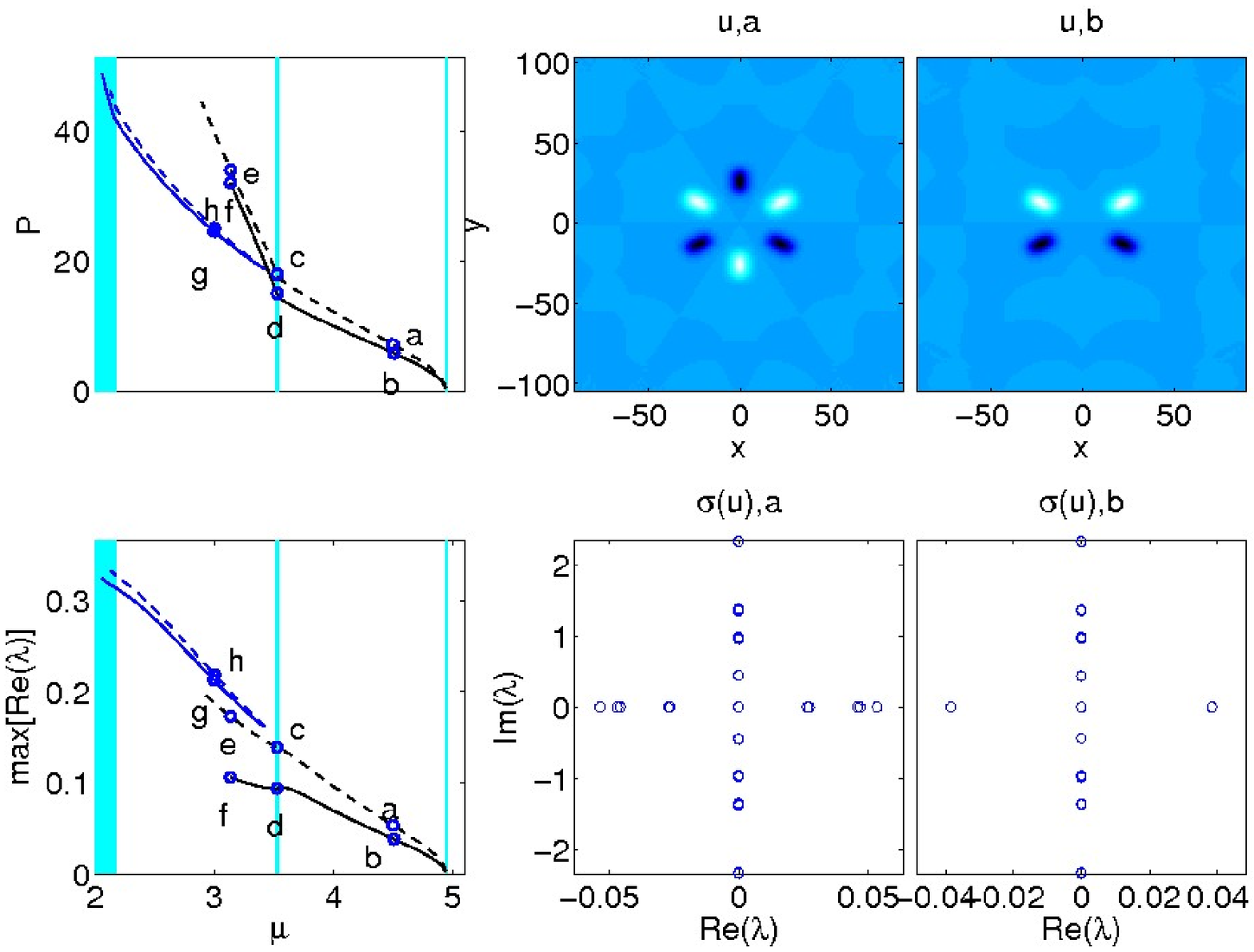}\\
\includegraphics[width=.5\textwidth]{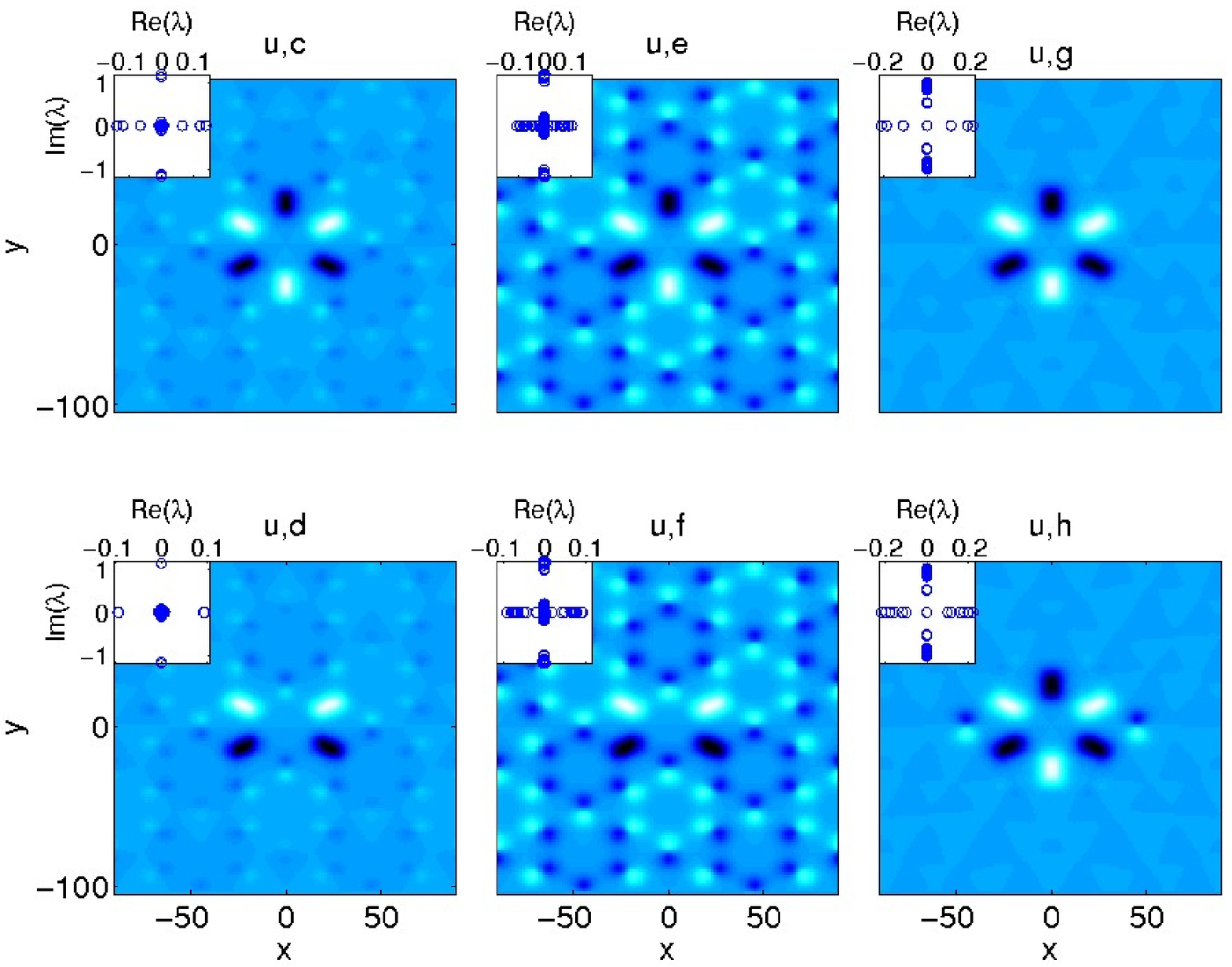}\\
\end{center}
\caption{(Color online)  The same panels as Fig. \ref{ip6c},
but for the out-of-phase hexapole.}
\label{op6c}
\end{figure}

%\begin{figure}[tbp!]
%\begin{center}
%\includegraphics[width=.55\textwidth]{fig10.eps.jpg.ps}
%\end{center}
%\caption{(Color online) The same panels as Fig. \ref{ip6dy} except
%for the unstable out-of phase solution given in Fig. \ref{op6c} (a).
%Again all six sites survive for a very long distance, although
%the phase correlation is lost very rapidly here due to the strong
%instability.}
%\label{op6dy}
%\end{figure}

\begin{figure}[tbp!]
\begin{center}
\includegraphics[width=.55\textwidth]{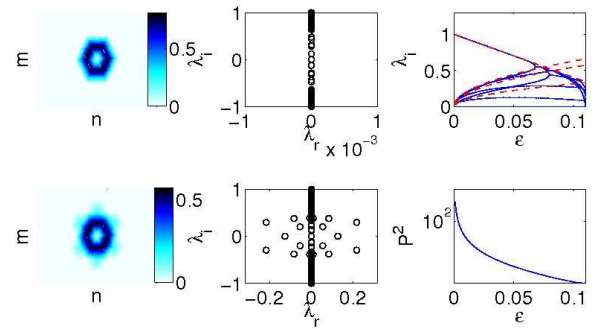}
\end{center}
\caption{(Color online) The same panels as Fig. \ref{ip6d}
except for the stable single charge vortex solution and
the modulus of the profiles are given, i.e. $|u|^2$ instead
of $u$.  The particular solutions are for $\varepsilon=0.06$ (top)
and $\varepsilon=0.11$ (bottom).}
\label{s1d}
\end{figure}

%\begin{figure}[tbp!]
%\begin{center}
%\includegraphics[width=.55\textwidth]{dyno_s1h.eps.jpg.ps}\\
%\includegraphics[width=.55\textwidth]{dyno_s1.eps.jpg.ps}
%\end{center}
%\caption{(Color online) The same as Fig. \ref{ip6di} for the
%single charge vortex solution for $\varepsilon=0.1$.
%The configuration breaks up very rapidly due to the instability
%and subsequently most of the power is concentrated on two
%sites opposite to each other which remain close in amplitude
%and phase for a long distance.}
%\label{s1di}
%\end{figure}

\begin{figure}[tbp!]
\begin{center}
\includegraphics[width=.5\textwidth]{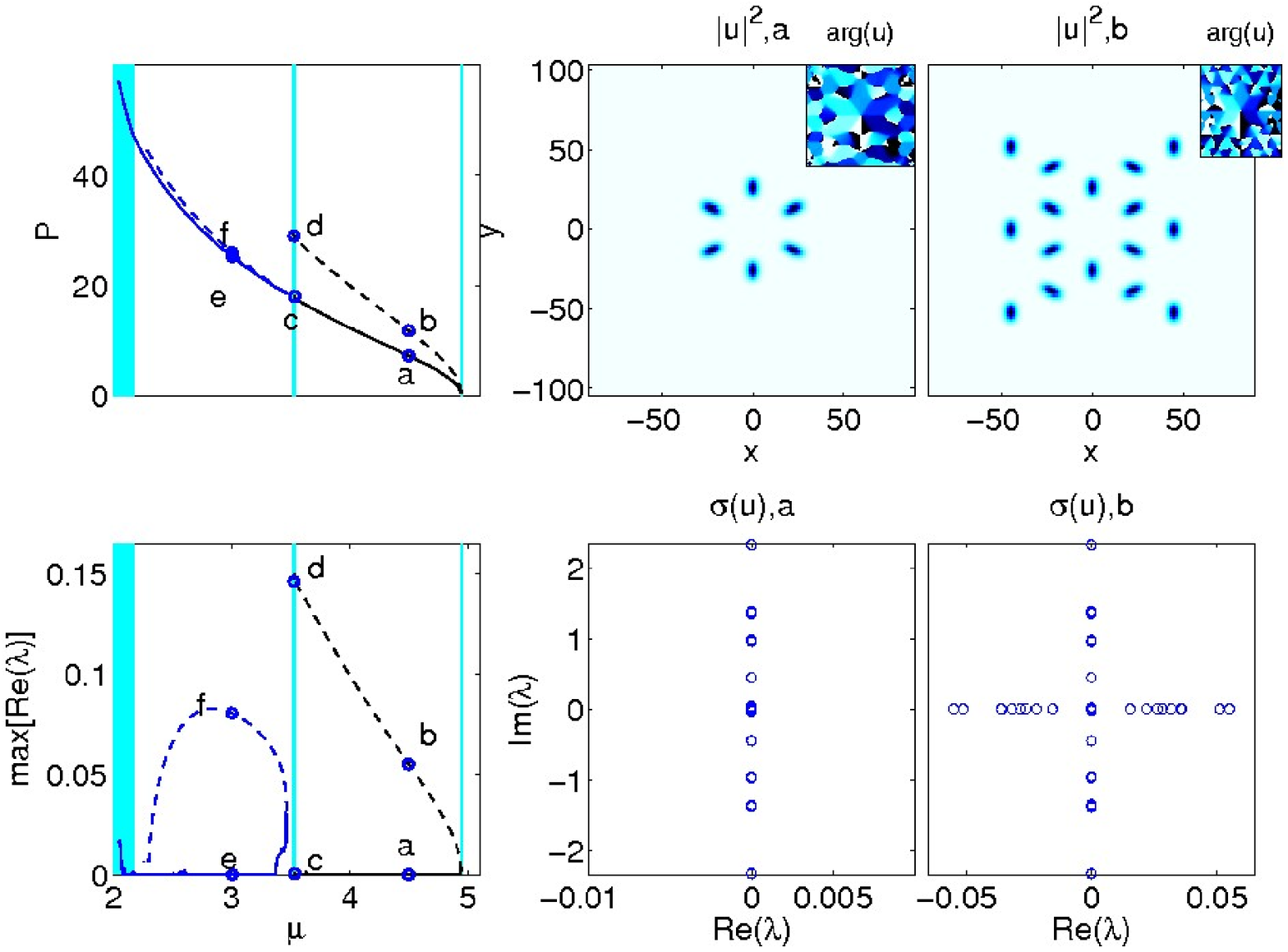}\\
\includegraphics[width=.5\textwidth]{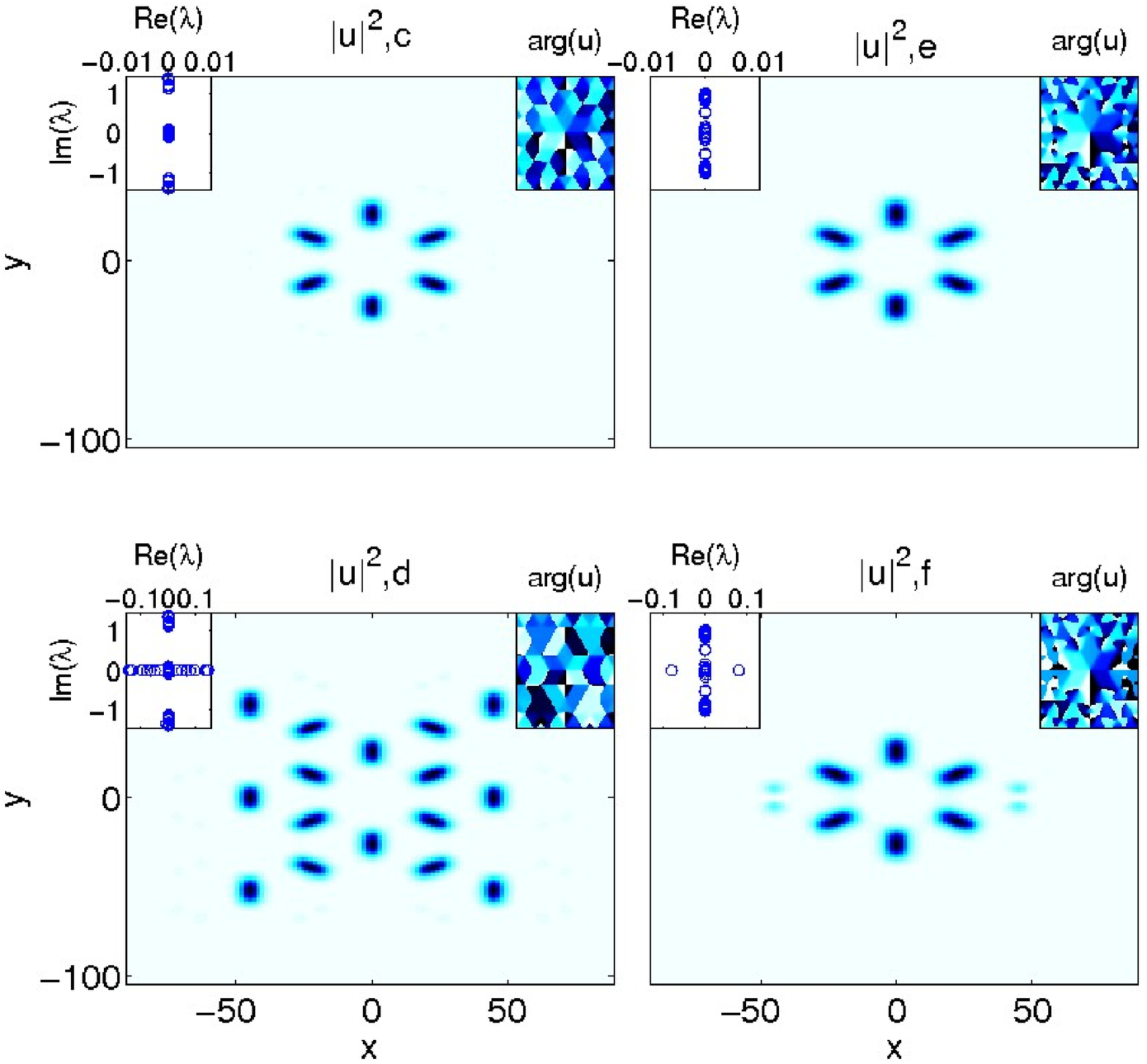}\\
\end{center}
\caption{(Color online) The same panels as Fig. \ref{ip6c},
but for the  single charge vortex solution and again, as in
the analogous discrete case, the modulus is given in lieu
of the field itself. Here there are
small embedded panels in the top right corners of the profile
images with the phase of the solution.
Also, we were not able to identify continuations of the
branches from the first gap into extended solutions in the
second gap in this case, although we did identify semi-extended
states.}
\label{s1c}
\end{figure}

%\begin{figure}[tbp!]
%\begin{center}
%\includegraphics[width=.55\textwidth]{s1_dynop.eps.jpg.ps}
%\end{center}
%\caption{(Color online) The same as Fig. \ref{ip6dy} but for the single
%charge vortex from Fig. \ref{s1c} (b).  One can see the phases maintain
%the correlation until $z=50$, although then it decomposes as the
%vortex leaves the center of the configuration.}
%\label{s1dy}
%\end{figure}

Stability is examined by
linearizing
Eq. (\ref{eq1}) and its conjugate
around an exact
stationary solution $(u,u^*)^T$
to Eq. (\ref{sta_eq}).
%for some coupling $\varepsilon$.
If we assume the perturbation
is separable of
the form $\tilde{u}=e^{\lambda t} w(x)$,
then we have the following
%linearizaed problem
eigenvalue problem for $i \lambda$
%and expanding the leading order
%perturbation into eigenfunctions and eigenvalues,
%we obtain
%is given by
%the following Bogoliubov system
%\begin{center}
%\begin{math}
\begin{equation}
\label{linstab}
\left\{  i\lambda I_2 + \bordermatrix{& \cr & I & 0\cr
 & 0 & -I \cr} \left [ \mathcal{J}(u)+
\mathcal{L}_2
%\bordermatrix{& \cr & I & 0\cr
% & 0 & I \cr}
\right ] \right \} v=0,
\end{equation}
%\end{math}
%\end{center}

\noindent where $\mathcal{J}$ is defined in Eq. (\ref{discrete_lin1})
and $\mathcal{L}_2, I_2$ are the $2\times2$ block diagonal matrices with
$\mathcal{L}, I$ (respectively) on the diagonal.
The eigenvalue problem imposes no restriction on the eigenvectors
$v$. This form of the Jacobian is identical for
the discrete and continuum versions of the problem,
up to the definitions of the operators $\mathcal{N}$
and $\mathcal{L}$ and the domain in which the spatial variables live.
The modest size of this matrix for the discrete
case, $[2 \times 33 \times 33]^2$ is not a problem for a
full diagonalization and we implement the MATLAB
function {\bf eig} to do so.  On the other hand,
the matrix for the model of the continuous domain
is $[2 \times 120 \times 120]^2$ and cannot be inverted.
Fortunately a standard finite difference discretization
leads to a sparse banded matrix which is perfectly
suited for Arnoldi iterative algorithms which minimize
memory and use successive
approximations of eigenvalues and eigenvectors until
convergence. Such a method is
implemented by the MATLAB function {\bf eigs}, which
we use here.

Twofold symmetry over each of the real and imaginary
axes is guaranteed by the fact that the Jacobian
of the full problem, which defines the linearization
at any point,
%is {\it Hamiltonian}
$H$ has the property that
%(i.e.
$JHJ=H^T$ (where $J$ is the canonical symplectic matrix
having the properties $JJ=-I$ and $JJ^T=I$,
which implies $e^H=M$ is {\it symplectic}
or $M^T JM=J$), and so
${\rm Re}(\lambda_j) \neq 0$ implies an
instability.

In the discrete case, the stability will be compared to
analytical results for small $\varepsilon$ based on
Lyapunov-Schmidt analysis of the expansion of the
equation around the AC limit (see, e.g. \cite{pgk_dnls} and
references therein for details).  
%Basically, the linearization
%around the solution to Eq. \ref{sta_eq} for nonzero $\varepsilon$ 
%can be assumed to live in the ${\rm ker}\{\mathcal{J}(u_0)\}=\{e_j\}$ 
%and, hence, the projection of this linearization 
%onto the kernel amounts to just a change of basis 
%and is equivalent to the linearization of the original problem 
%in the changed basis, or, the persistence condition 
%given in Eq. \ref{persist}.
For the contour M, there are $|M|$
%the stability can be determined from the
eigenvalues $\gamma_j$ of the $|M| \times |M|$
Jacobian ${\cal M}_{jk}=\partial g_j/\partial \theta_k$ of
the diffeomorphism given in Eq. (\ref{persist}).
Now, for each excited site, there are a zero and 
a negative eigenvalue of $\{\mathcal{J}(u_0)\}$, which are both 
mapped to zero eigenvalues of $\bordermatrix{& \cr & I & 0\cr
& 0 & -I \cr}\{\mathcal{J}(u_0)\}$.  
So, one can follow the same procedure for the 
eigenvalue problem of the linearization (Eq. \ref{linstab})
of the original problem, except projecting also onto the
generalized kernel and expanding to lower order 
($\sqrt{\varepsilon}$) and, hence, obtain a mapping 
between its eigenvalues and those of the Jacobian of ${\bf g}$. 
In particular, for each eigenvalue $\gamma_j$ of ${\cal M}$, 
the full linearization
around a stationary solution with non-zero nodes
in $M$, given by Eq. (\ref{linstab}),
will have an eigenvalue pair $\lambda_j$ given, to leading order, by
$\lambda_j = \pm \sqrt{2 \gamma_j \varepsilon}$ in the case that the
sites in $M$ are nearest neighbors of each other.
If the excited nodes are separated by a site
then, for the DNLS model eigenvalues, $\varepsilon$ is replaced
by $\varepsilon^2$ in the previous relation.
The Jacobian matrix ${\cal M}$ has the following form:

\begin{eqnarray}
\label{Melements3}
\begin{array}{lcl}
({\cal M})_{j,k} = & \\
 \left\{ \begin{array}{lcl}
-\cos(\theta_{j+1} - \theta_j) - \cos(\theta_{j-1} - \theta_j), &
\quad & j = k, \\  \cos(\theta_j - \theta_k), & \quad & j = k \pm 1, \\
0, & \quad & |k - j | \geq 2. \end{array} \right.
\end{array}
\end{eqnarray}

We now briefly discuss the principal stability conclusions, for
the defocusing case of \cite{ourpre,ourhon}, which we expect
to remain valid in the present configuration.
Nearest neighbor excitations in the defocusing case correspond
to nearest neighbor excitations in the focusing case, but
with an additional $\pi$ phase in the relative phase of the
sites added by the so-called staggering transformation \cite{ourpre}.
Therefore, the in-phase nearest neighbor configuration in the defocusing
case corresponds to an out-of-phase such configuration in the focusing case
(and should thus be stable) \cite{pgk_dnls}. On the other hand,
next nearest neighbor out-of-phase defocusing configurations
would correspond
to next nearest neighbor out-of-phase focusing configurations and
should also be stable (at least in some parameter regimes). By the
same token, out-of-phase nearest neighbor, and
in-phase next nearest neighbor
structures should be unstable.
%These considerations also indicate
%that in-phase opposite dipoles should be stable, while out-of-phase
%such dipoles should always be unstable.
Finally, the single-charge vortex
%structures
and in-phase hexapoles should be stable,
while the double charge vortex and out-of-phase hexapoles
should be unstable.
However, notice that, as discussed in \cite{ourpre},
the multipole structures
characterized as potentially stable above will, in fact, typically possess
imaginary eigenvalues of negative Krein signature (see e.g. \cite{kks}
and references therein). These may lead to oscillatory instabilities
through complex quartets of eigenvalues, which arise by means
of Hamiltonian-Hopf (HH)
bifurcations \cite{vdm} emerging from collisions with eigenvalues
of opposite (i.e., positive) Krein signature. These conclusions
will be discussed in connection with our detailed numerical results
in what follows.

\section{Numerical results}
\label{numerics}

Now, the above theoretical predictions will be matched
against systematic numerical simulations.  First, for the
discrete model, we will perform continuations
in the coupling parameter from the AC limit in order to
compare the resulting relevant eigenvalues from the
linearization spectrum with the corresponding prediction.  Next, we will
test these results against the continuum model.  The
stability results from the discrete case are expected to
hold in the sense that there will either be real
eigenvalue pairs in the spectrum of the solutions whose
discrete analog is unstable close to the AC limit, or else
there will be intervals of stability and quartets of
eigenvalues due to HH and inverse HH bifurcations. The
continuum model reveals not only gap soliton solutions
in both the first and the second gaps, but also 
solitons from where the branch of solutions
from the first gap passes through the first band and
subsequently becomes extended.
Unstable solutions were evolved in time in order to
observe their dynamical behavior.  Most solutions in the
discrete case decompose into breathing configurations with fewer
populated sites and some interesting phase correlations.  In the
continuum case, most configurations survive for a long propagation distance,
with instabilities manifested only as phase reshaping.

This section will be composed of two
parts, the first of which will address configurations with
six neighbors on the hexagonal cell, the results of which
are consistent with recent results in the continuum honeycomb
defocusing case \cite{ourhon} and
both honeycomb and hexagonal \cite{ourhexhon} focusing cases
(translated with the appropriate staggering transform
along the contour).  In the second part we will look at
quadrupoles along the four corners of the ``hourglass"
cell which is unique to the Kagom{\'e} lattice.

\subsection{Vortices and hexapoles in the hexagonal cell}

The results for the six-site configurations in the hexagonal
cell are presented in this section.  First we will consider
the real-valued configurations of $\Delta\theta=0$ and
$\Delta\theta=\pi$ and then the complex-valued ones where
$\Delta\theta=\pi/3$ and $\Delta\theta=2 \pi/3$.

\subsubsection{In-phase}

Here we will consider the results of the predictions from the previous
section for the six-site in-phase configuration (i.e. $\Delta\theta=0$).
This configuration has been predicted to be {\it stable}.  In the discrete
model close to the AC limit, Eq. \ref{Melements3} predicts, to first
order in $\varepsilon$, two double
pairs of eigenvalues $i\sqrt{2 \varepsilon},i\sqrt{6 \varepsilon}$ and
single pairs at $i\sqrt{8 \varepsilon}$ and $0$.
Here we digress slightly to discuss the bound of the phonon band.
We consider plane waves of the form $w=e^{i(p n +q m)}$, in each
of 2 (of the 3) principal directions, which are used to index the 2-dimensional
lattice \cite{kouk}, $(m,n)$.  Since there are 3 types of nodes
in this case, each having neighbors in 2 of the 3 principal directions
(that are the same for the hexagoinal lattice),
we must consider a linear combination of equal $1/3$ weights
of the corresponding dispersion relation for each type. 
This is equivalent to $2/3$ of the dispersion relation of the 
hexagonal lattice, i.e.

\begin{equation}
\mathcal{L}_{\varepsilon}w=(4-\frac{4}{3}[\cos(q)+\cos(p)+\cos(p+q)])
\varepsilon< 6\varepsilon.
\end{equation}

So, the smallest eigenvalue of the phonon band is given by
$i(1-6\varepsilon)$, and upon its collision with the eigenvalues
which bifurcated from the origin, a cascade of HH bifurcations ensues.
The numerical results are presented in Fig.
\ref{ip6d}.  The left column displays two solutions, before (top)
 and after (bottom), the continuous spectrum intersects with the
bifurcation eigenvalues.  The middle column has the corresponding
linearization spectra.  The top right panel depicts the imaginary
component of the bifurcation eigenvalues as given numerically (solid
line) and by the first order theoretical approximation (dashed line).

The dynamical
evolution of the unstable solution from the bottom row is displayed in
Fig. \ref{ip6di}.  Eventually the instability manifests 
itself and the original
configuration is destroyed. 
Two nearest-neighbor sites remain with different
amplitudes that oscillate.  When they are closer in amplitude
they are out-of-phase
while when they are further apart, they are in-phase.  It is worth noting
here that a similar phenomenon was found in the hexagonal as well as the
honeycomb lattices with a focusing nonlinearity in 
\cite{ourhexhon}, except with
the relative phases reversed, i.e. the sites were in-phase when closer
and out-of-phase when further apart, presumably due to the nature
of the nonlinearity (focusing versus the defocusing one here).  
%It is in-phase neighbors that
%are stable here and out-of-phase that are unstable, so one may hypothesize
%that when the phases match, the dynamical state
%attempts to adjust to a uniform amplitude stable
%in-phase dipole and the amplitudes become closer, and then, when the
%phases become opposite, it readjusts due to the repulsion from the
%opposite phase dipole state, 
%%with the amplitudes becoming
%%more uneven again, 
%and the cycle repeats.

Next we investigate the in-phase hexapole in the continuum setting.
The solution is stable in the entire first band-gap
(see Fig. \ref{ip6c}, top two rows and $a$).  When it reaches
the first band it collides with an unstable branch (b) which has two
neighboring wells populated out-of-phase and disappears in a saddle-node
bifurcation.  When these branches reach the second band, they 
immediately become unstable
as they reshape into extended solutions (see e,f).  There does exist
a second-gap soliton solution which is stable, (g).  This solution
%bifurcates
disappears in a saddle-node bifurcation with a marginally stable solution
that has next-nearest-neighbor wells on four
sides populated with intra-site dipoles.  The evolution of solution (b)
is represented in Fig. \ref{ip6dy} by the phase of the sites via the
following quantity
${\rm arg}(u) \chi \{(x,y) | |u|^2>0.5 {\rm max}_{(x,y)}[|u|^2] \}$,
where $\chi$ is the indicator function of the set that annihilates
%suppresses
the field outside that set.
All sites remain for a long propagation 
distance, although the phase correlation is
lost by $z=300$.

\subsubsection{Out-of-phase}

In this section, we consider the {\it unstable} out-of-phase hexapole
 ($\Delta \theta =\pi$).
First, in the
discrete case presented in Fig. \ref{op6d} the theoretical prediction
of linearization eigenvalues, which are exactly a factor $-i$ times
those for the in-phase solution, are
confirmed for small $\varepsilon$.  The dashed line $1-6 \varepsilon$,
which represents the smallest phonon eigenvalue, is included
here to show that the actual linearization eigenvalues remain bounded by
this line, similarly to what was observed for square lattices in 
\cite{ourpre}. 
%even though they are real-valued and there should be no
%a priori correlation.  
A solution is shown for small coupling and large
gap, as well as one when the gap is closing and the solution decaying.
The dynamics in Fig. \ref{op6di} reveal that all
sites survive for a long propagation distance, and,
while there is no clear phase correlation between all sites, there is
between some.  For instance, the largest amplitude two
next-nearest-neighbor lobes remain out-of-phase.  This is again
consistent with a feature that was recently observed in \cite{ourhexhon},
since next-nearest-neighbor interactions are expected to be the same
for focusing and defocusing non-linearities due to the staggering
transformation \cite{pgk_dnls}.  The middle amplitude site next-nearest to
both of these oscillates between in-phase with one and then the other,
while there is no apparent correlation of the other three smaller
amplitude sites.

The same panels as the in-phase case, Fig. \ref{ip6c}, are shown for
the continuum version of the out-of-phase hexapole in Fig. \ref{op6c}.
The solution this time actually collides with a
%slightly more stable
four-well structure, which is slightly more stable, due to fewer
unstable pairs of populated wells.  Again there are continuations
through the second band to extended states and again there exists
a disjoint branch of second gap states.  All states here are unstable.
Under dynamical evolution, 
%is presented again by the phase because 
again all six
sites remain for a long propagation 
distance (not shown).
%(Fig. \ref{op6dy}).  
However, the phase
correlation breaks down as early as $z=20$, due to the instability.

\subsubsection{Single charge vortices}

Next we look at the {\it stable} single charge vortex solution
($\Delta \theta=\pi/3$).
The discrete problem
is predicted to be stable with double pairs of eigenvalues at
$\pm i \sqrt{\varepsilon}$ and $\pm i \sqrt{3 \varepsilon}$, and single pairs
at $\pm2 i \sqrt{\varepsilon}$ and $0$.  The prediction is confirmed
in Fig. \ref{s1d} and there is good agreement until the HH bifurcations
set in with the continuous
spectrum when $\sqrt{\varepsilon}\approx \lambda_i=1-6\varepsilon$.
A cascade of such bifurcations
follows and an example profile and spectrum after this time are shown
in the bottom left.  The evolution of the unstable solution from the
bottom row was investigated (not shown) and
%is given in Fig. \ref{s1di}.  F
four sites decompose into
essentially background radiation, while two cells opposite to one another
inherit most of the power and remain close in amplitude and in-phase
for a long distance.  This is again in
agreement with the results of \cite{ourhexhon}, since in-phase opposite
sites on the honeycomb cell are
next-to-next-nearest-neighbors, so with a defocusing nonlinearity
it is equivalent to an out-of-phase
pair with a focusing nonlinearity.  Many comparable amplitude
out-of-phase breathers were found in \cite{ourhexhon}.

The continuum version of this configuration given in Fig. \ref{s1c} (a)
collides in a saddle-node
bifurcation with an unstable configuration with additional populated
sites on the perimeter, (b). The second gap version (e) collides
with an unstable
state (f) having two intrasite dipoles populated outside the original vortex.
However, this state appears to stabilize closer to the third band.  
In the dynamical evolution of (b, not shown) 
%are presented in Fig. \ref{s1dy} where 
again the original
configuration of ``mass" survives for a long propagation
distance, while the phase
correlation decomposes by $z=100$. 
%Fig. \ref{s1c} (b).

\subsubsection{Double-charge vortices}

The last of the six-site configurations we consider is the double-charge
vortex ($\Delta \theta =2 \pi/3$).
%The discrete results are presented in Fig. \ref{s2d}.  
The
stability predictions are again confirmed for small $\varepsilon$;
however, due to the instability of the branch throughout its
existence range, we do not present the details here for the sake
of brevity.  
%The dynamics of
%the solution from the top row are given in Fig. \ref{s2di} and one can
%notice another similar configuration to those from \cite{ourhexhon}; that
%is, two pairs of nearest-neighbor, uneven amplitude breathers oscillating
%with the pattern of the ones in Fig. \ref{ip6di}.
The continuum model admits similar branch structure for this 
solution as for previous cases. Here also, the solution is unstable
for all the cases examined. Its dynamics showed the principal
sites surviving for long propagation distances, although the structure
eventually disintigates. Again due to the generic instability of
this branch, numerical details are omitted here.
%,
%although due to convergence issues the entire bifurcation
%structure was not pursued.  All
%solutions are unstable (see Fig. \ref{s2c}).  The dynamics are similar
%in the sense that all sites survive for long distance (Fig. 18). The
%relative phase disruption only barely begins to occur by $z=100$.

\begin{figure}[tbp!]
\begin{center}
\includegraphics[width=.5\textwidth]{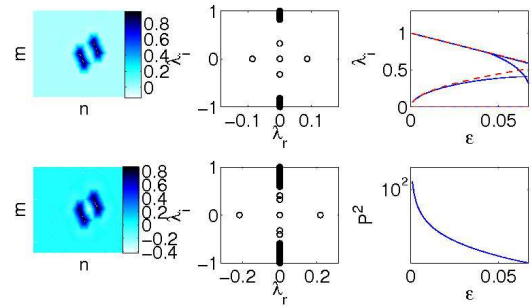}
\end{center}
\caption{(Color online) The same panels in the left two columns
as Fig. \ref{ip6d}
except for the unstable in-phase quadrupole in the hourglass
cell.
%Now there are real and imaginary eigenvalues, so the
%panels in the right column show them on the top and bottom,
%respectively.  The eigenvalues bifurcate in this case as they
%would if the opposite corners of the hourglass did not
%communicate.
The particular solutions shown are for $\varepsilon=0.03$
(top) and $\varepsilon=0.067$ (bottom).}
\label{ip4d}
\end{figure}

\begin{figure}[tbp!]
\begin{center}
\includegraphics[width=.55\textwidth]{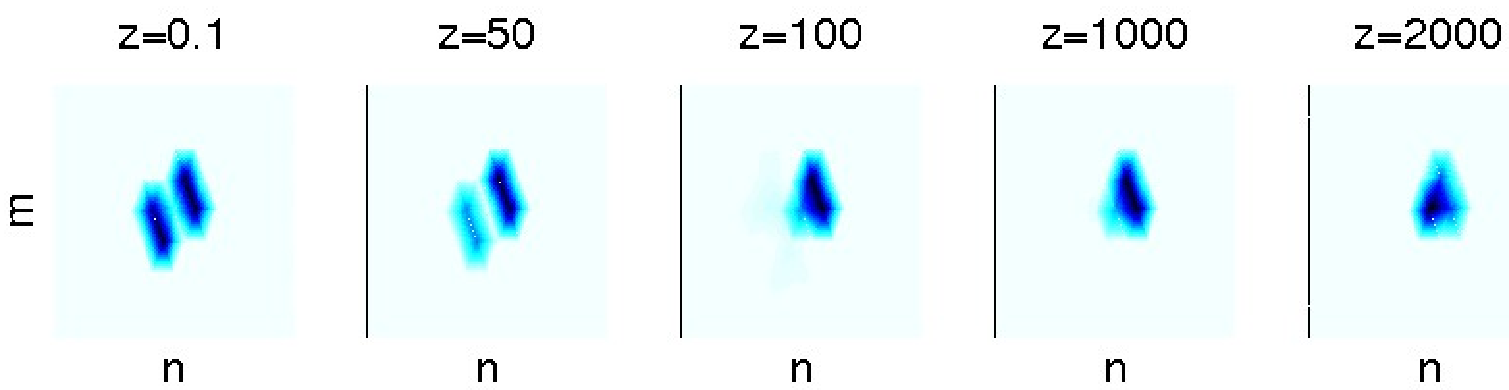}\\
\includegraphics[width=.55\textwidth]{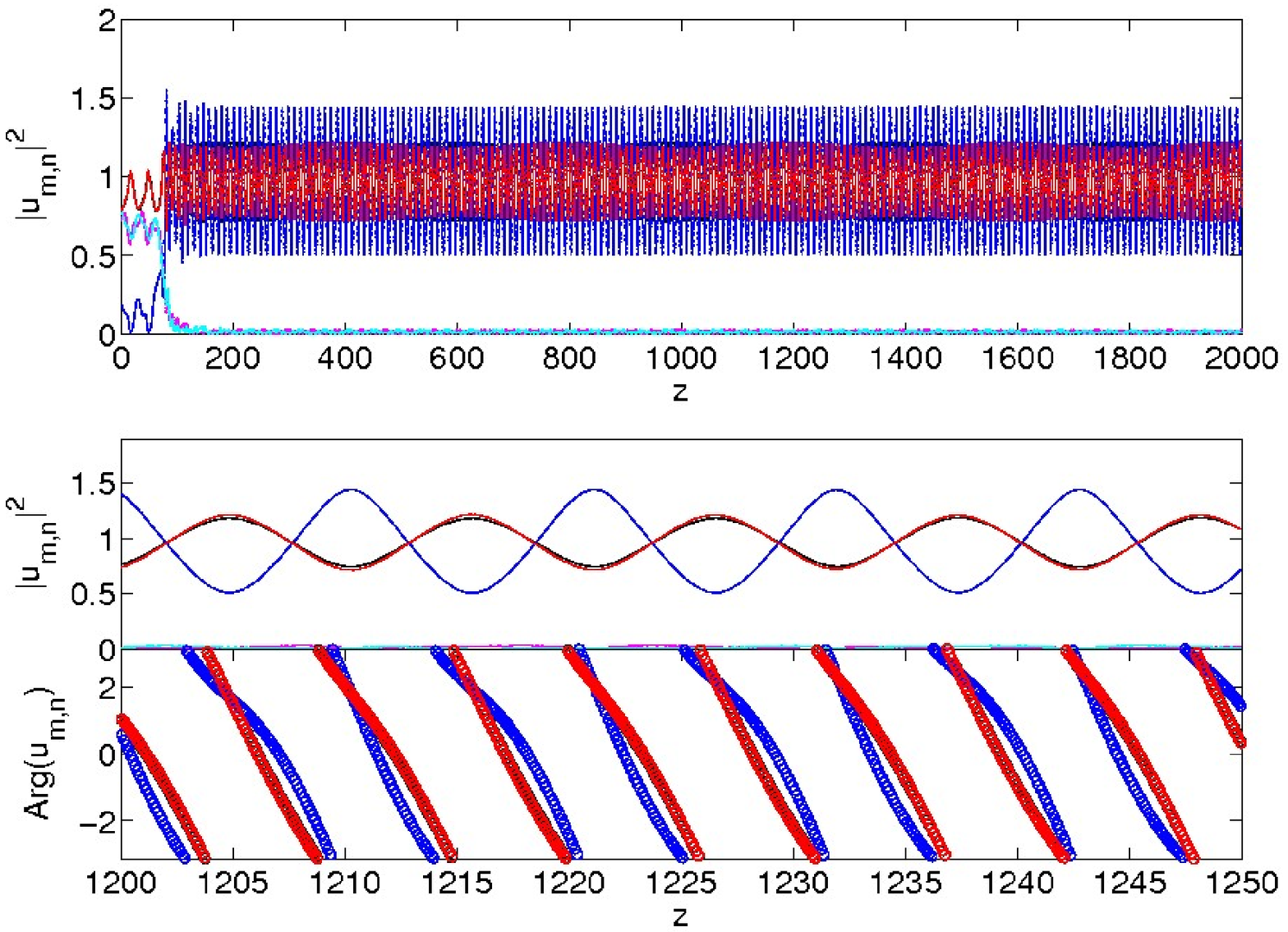}
\end{center}
\caption{(Color online) The same as Fig. \ref{ip6di} for the
in-phase quadrupole given in the bottom row
of Fig. \ref{ip4d}.  The center site becomes populated around $z=100$
and the remaining in-phase tripole (expected to be stable) persists
for a long distance with two sites almost exactly in-phase, and with equal
amplitudes, while the other has much larger oscillations in amplitude
opposite to the other two, but has very similar phase.}
\label{ip4di}
\end{figure}

\begin{figure}[tbp!]
\begin{center}
\includegraphics[width=.5\textwidth]{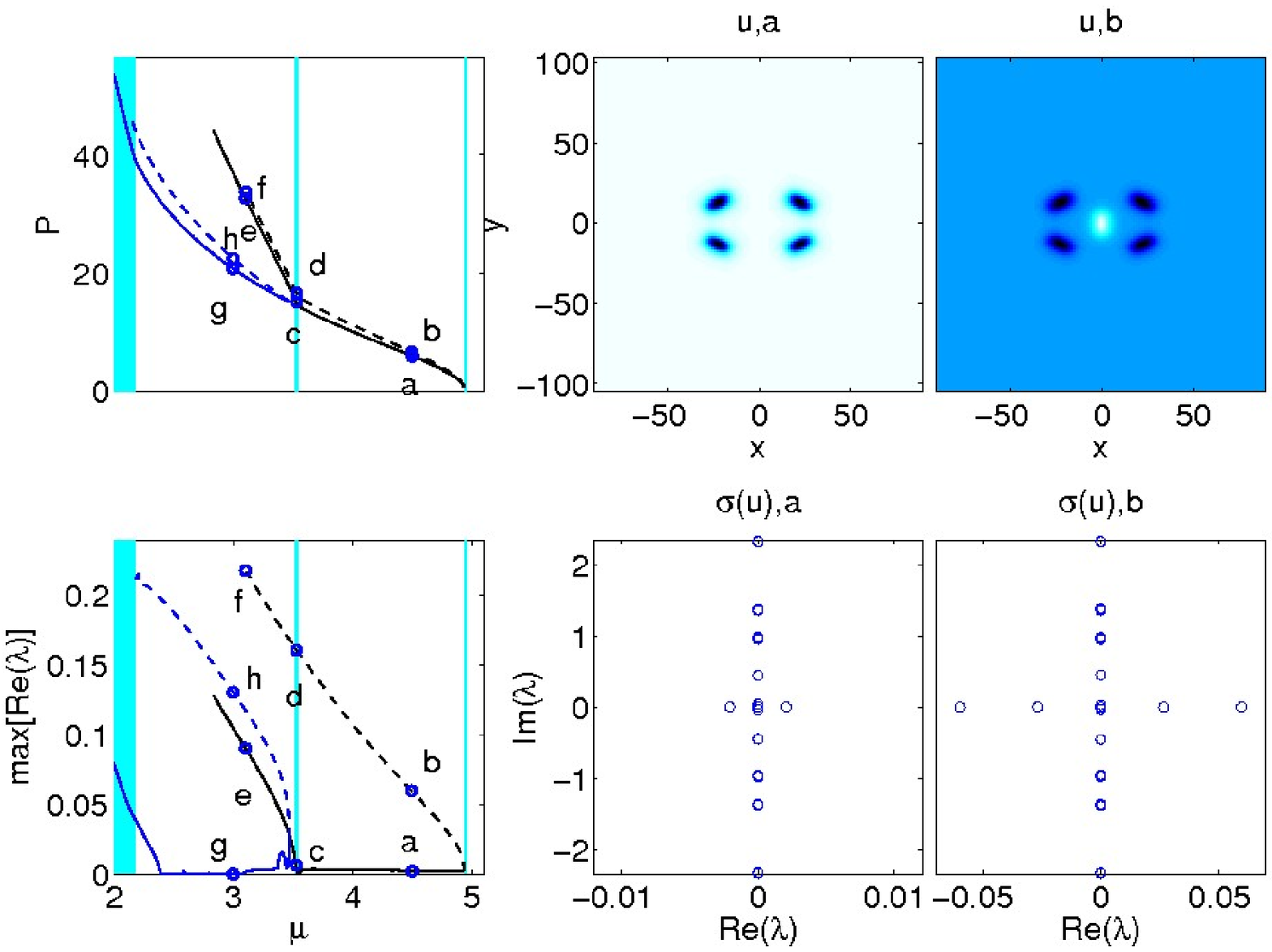}\\
\includegraphics[width=.5\textwidth]{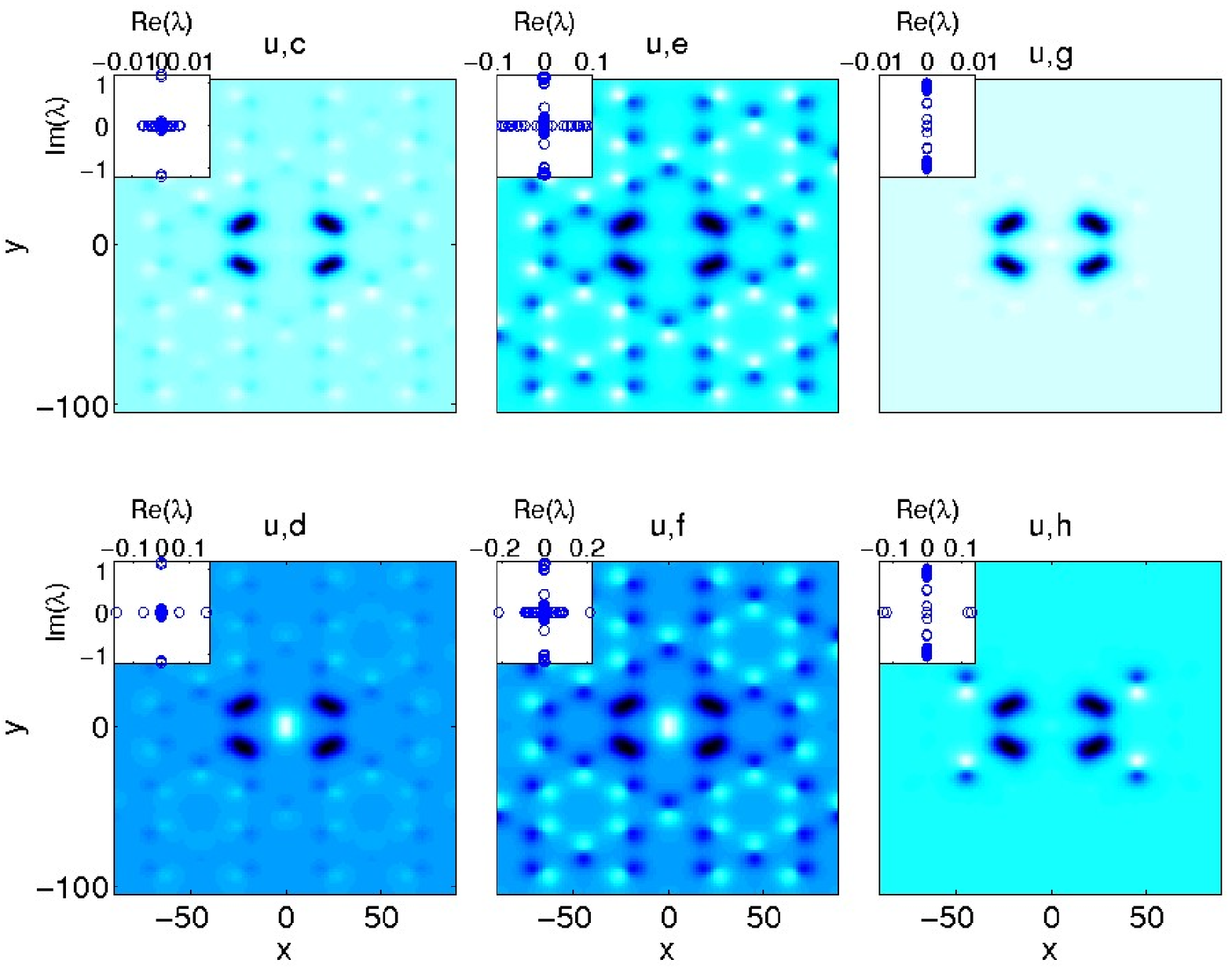}\\
\end{center}
\caption{(Color online) The same panels as Fig. \ref{ip6c},
but for the in-phase quadrupole.  A more unstable configuration
with the center well populated out-of-phase collides with this one
close to the first band-edge.}
\label{ip4c}
\end{figure}

\begin{figure}[tbp!]
\begin{center}
\includegraphics[width=.55\textwidth]{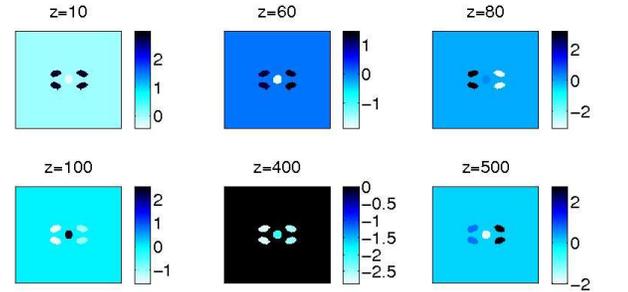}
\end{center}
\caption{(Color online) The same as Fig. \ref{ip6dy} but for the
in-phase quadrupole
from Fig. \ref{ip4c} (b).  The initial relative
phase loses the correlation after $z=60$ and
%center out-of-phase soliton
%quickly decays and the remaining 
%in-phase quadrupole maintains
%phase correlation 
the structure again persists for a long distance.}
\label{ip4dy}
\end{figure}

\begin{figure}[tbp!]
\begin{center}
\includegraphics[width=.5\textwidth]{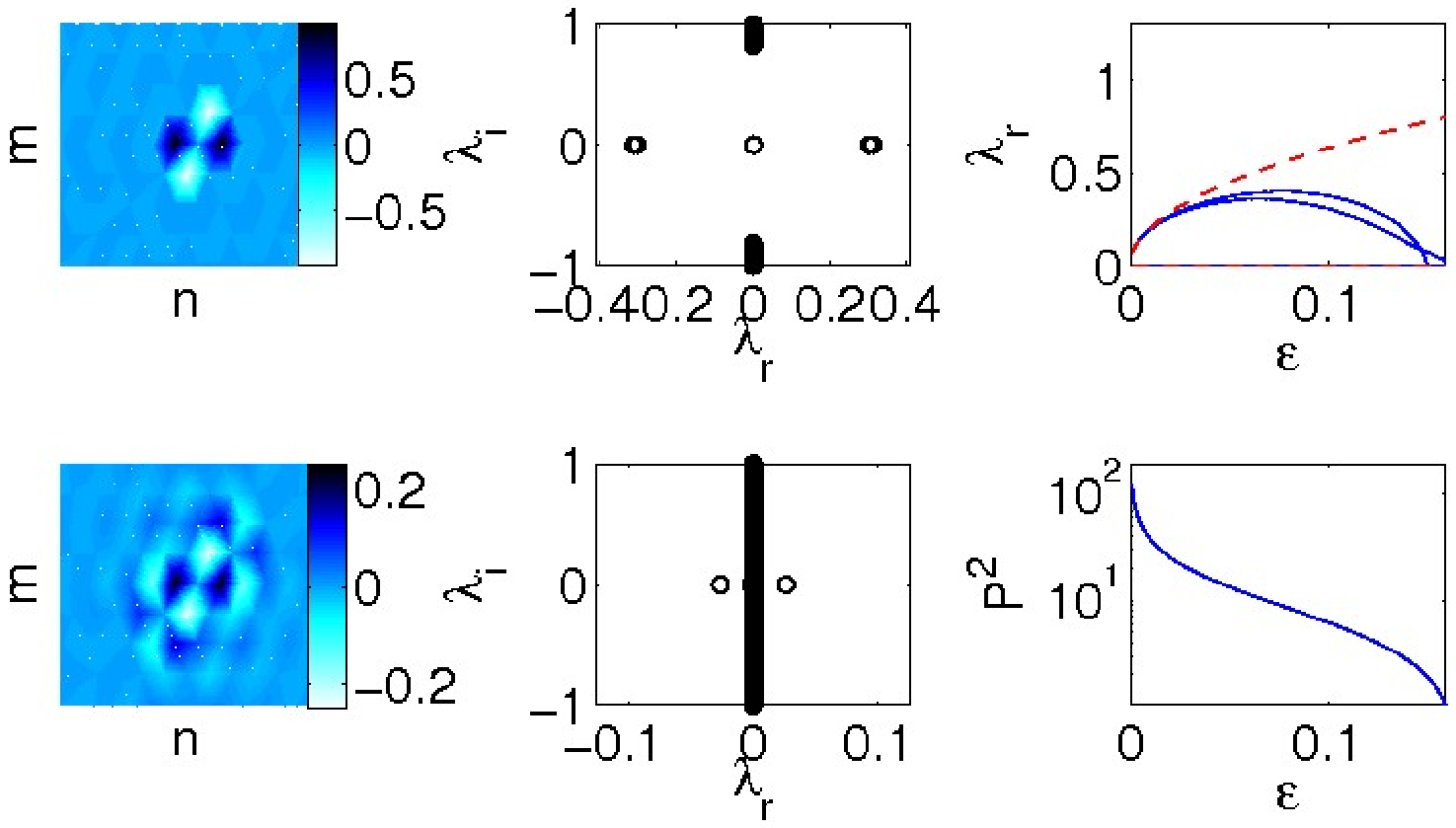}
\end{center}
\caption{(Color online) The same panels as Fig. \ref{ip4d}
except for the unstable out-of-phase quadrupole.
%Now the
%assumption that opposite corners do not communicate no longer
%holds and the imaginary eigenvalue predicted that way
%does not exist.  One does exist, but with very small amplitude,
%perhaps overpowered by the instability of the opposite corner
%next-nearest-neighbors.
The particular solutions shown are for $\varepsilon=0.03$
(top) and $\varepsilon=0.16$ (bottom).}
\label{op4d}
\end{figure}

%\begin{figure}[tbp!]
%\begin{center}
%\includegraphics[width=.55\textwidth]{dyno_op4h.eps.jpg.ps}\\
%\includegraphics[width=.55\textwidth]{dyno_op4.eps.jpg.ps}
%\end{center}
%\caption{(Color online) The same as Fig. \ref{ip6di} for the
%out-of-phase quadrupole solution given in the top row
%of Fig. \ref{op4d}. Again there are two pairs
%of uneven amplitude breathers with phases and amplitudes oscillating
%opposite to each other as in Fig. \ref{ip6d}.}
%\label{op4di}
%\end{figure}

\begin{figure}[tbp!]
\begin{center}
\includegraphics[width=.5\textwidth]{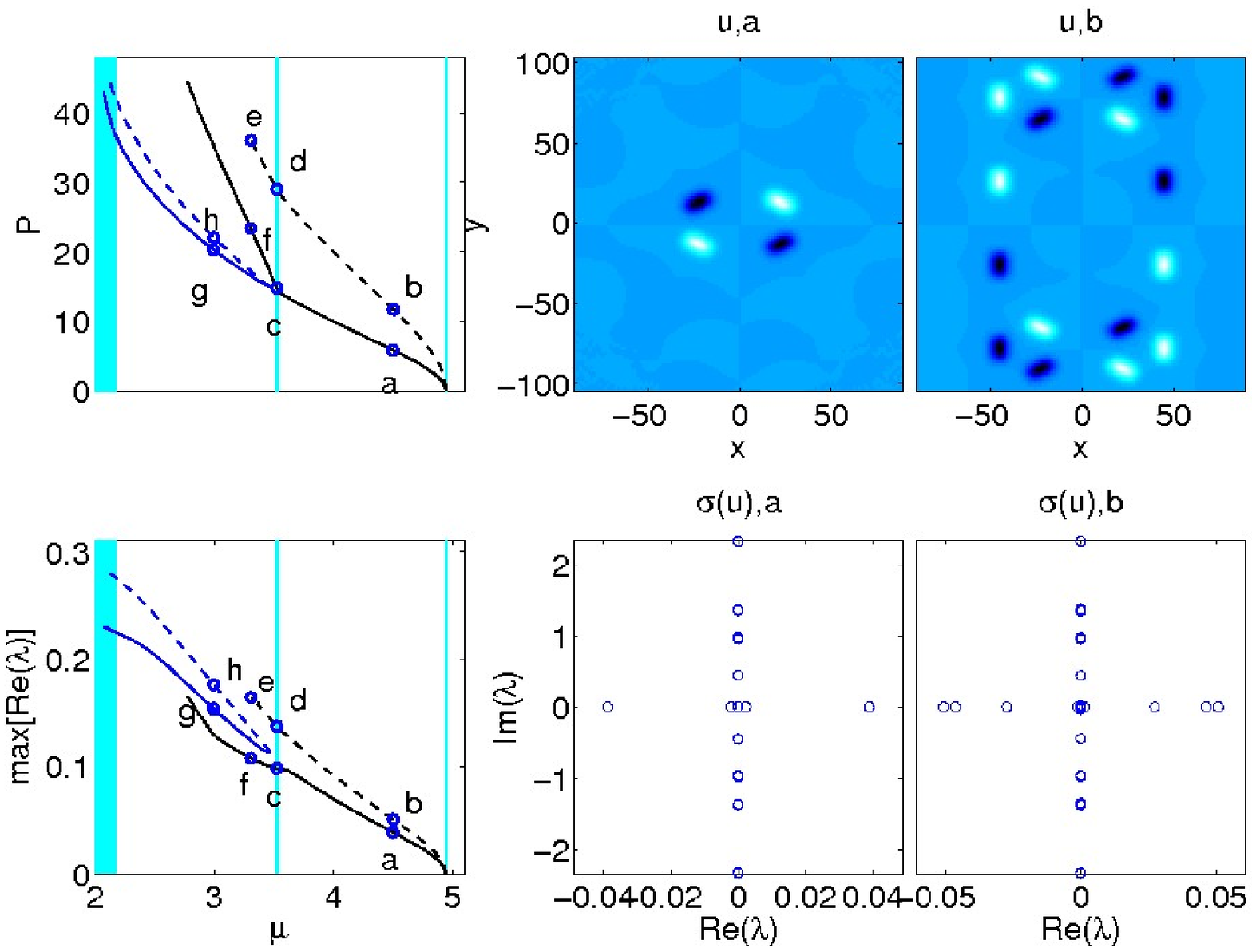}\\
\includegraphics[width=.5\textwidth]{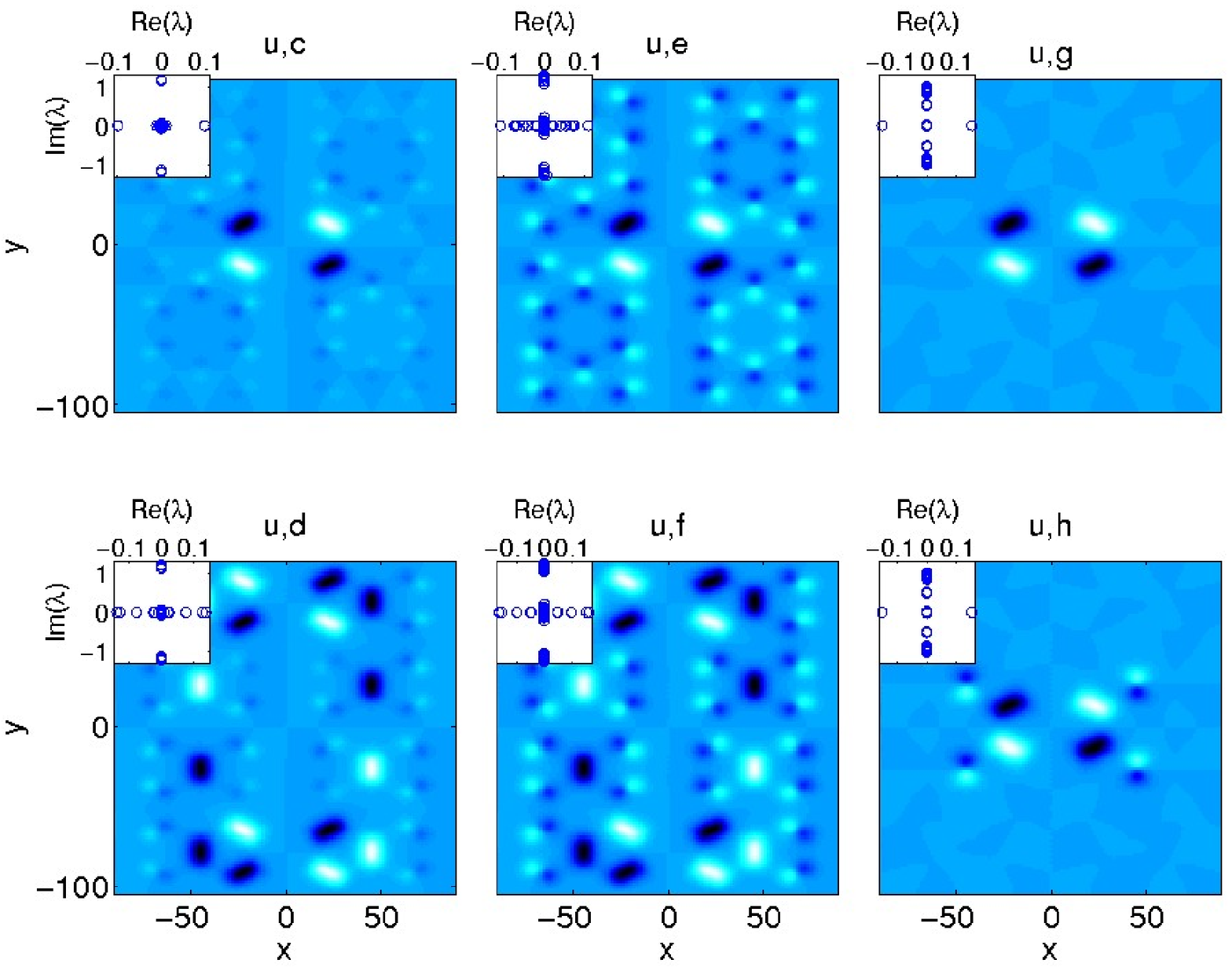}\\
\end{center}
\caption{(Color online) The same panels as Fig. \ref{ip6c},
but for the out-of-phase quadrupole. It collides with 
a branch that has a similar phase pattern as the original configuration.}
%an
%unexpected branch which does not have the original configuration
%as a subset, but has the same phase.}
\label{op4c}
\end{figure}

\begin{figure}[tbp!]
\begin{center}
\includegraphics[width=.55\textwidth]{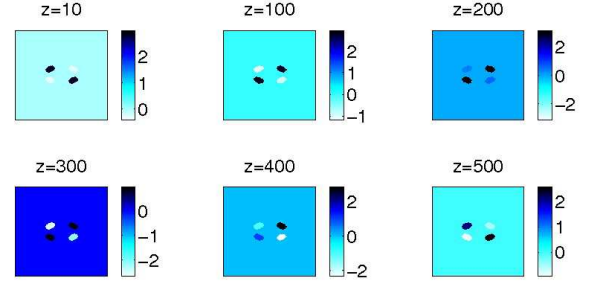}
\end{center}
\caption{(Color online) The same as Fig. \ref{ip6dy} but for
the unstable out-of-phase quadrupole given in Fig. \ref{op4c} (a).}
\label{op4dy}
\end{figure}

\begin{figure}[tbp!]
\begin{center}
\includegraphics[width=.5\textwidth]{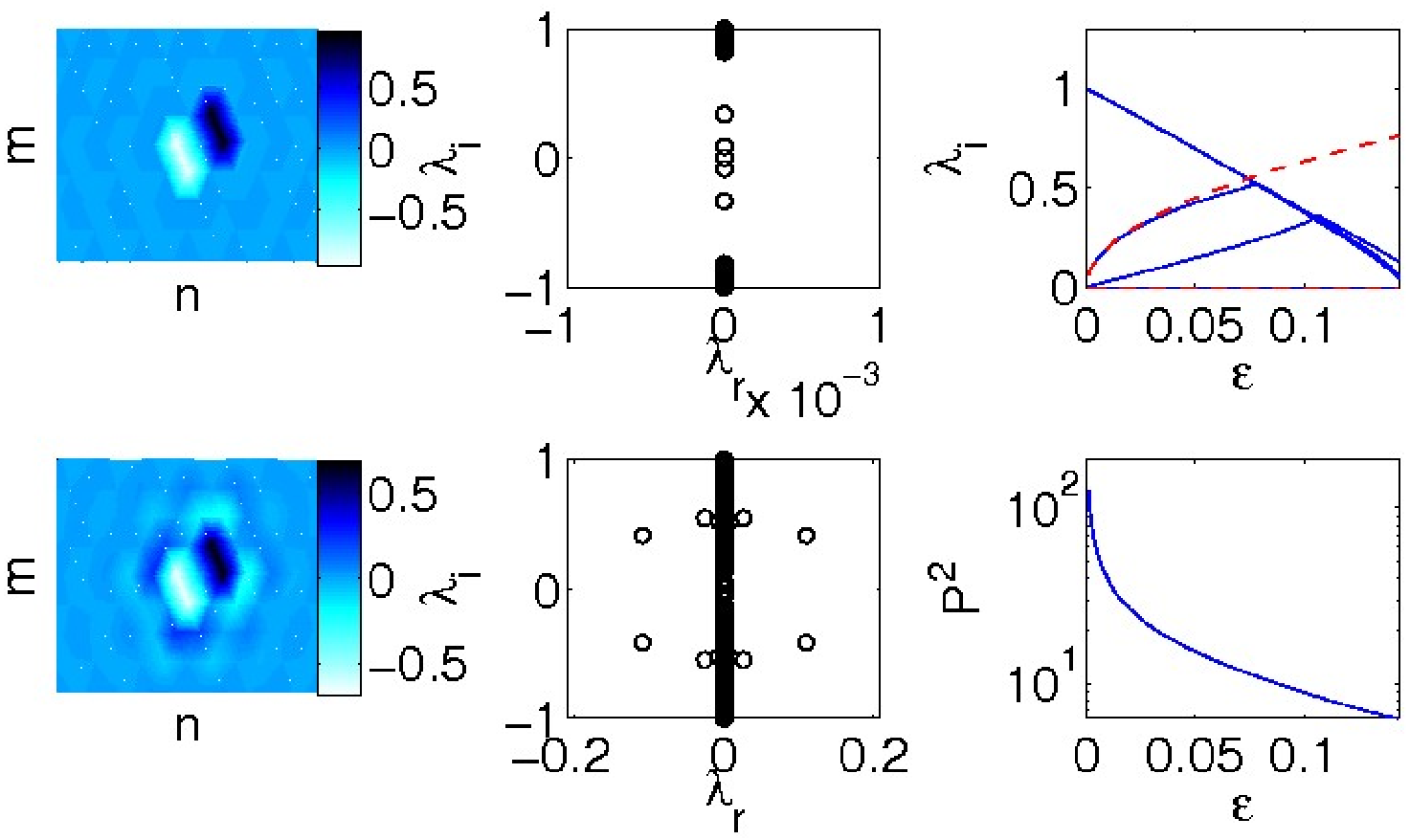}
\end{center}
\caption{(Color online) The same panels as Fig. \ref{ip6d}
except for the stable in-phase/out-of-phase quadrupole.
 The particular solutions shown are for $\varepsilon=0.03$
(top) and $\varepsilon=0.144$ (bottom).}
\label{ipop4d}
\end{figure}

%\begin{figure}[tbp!]
%\begin{center}
%\includegraphics[width=.55\textwidth]{dyno_ipop4h.eps.jpg.ps}\\
%\includegraphics[width=.55\textwidth]{dyno_ipop4.eps.jpg.ps}
%\end{center}
%\caption{(Color online) The same as Fig. \ref{ip6di} for the
%in-phase/out-of-phase solution given in the bottom row
%of Fig. \ref{ipop4d}.}
%\label{ipop4di}
%\end{figure}

\begin{figure}[tbp!]
\begin{center}
\includegraphics[width=.5\textwidth]{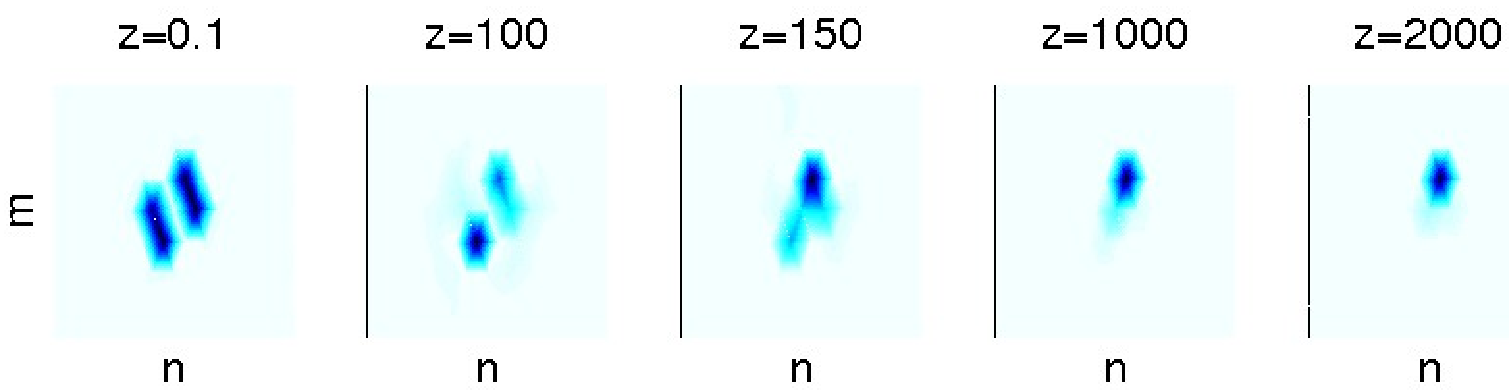}\\
\includegraphics[width=.5\textwidth]{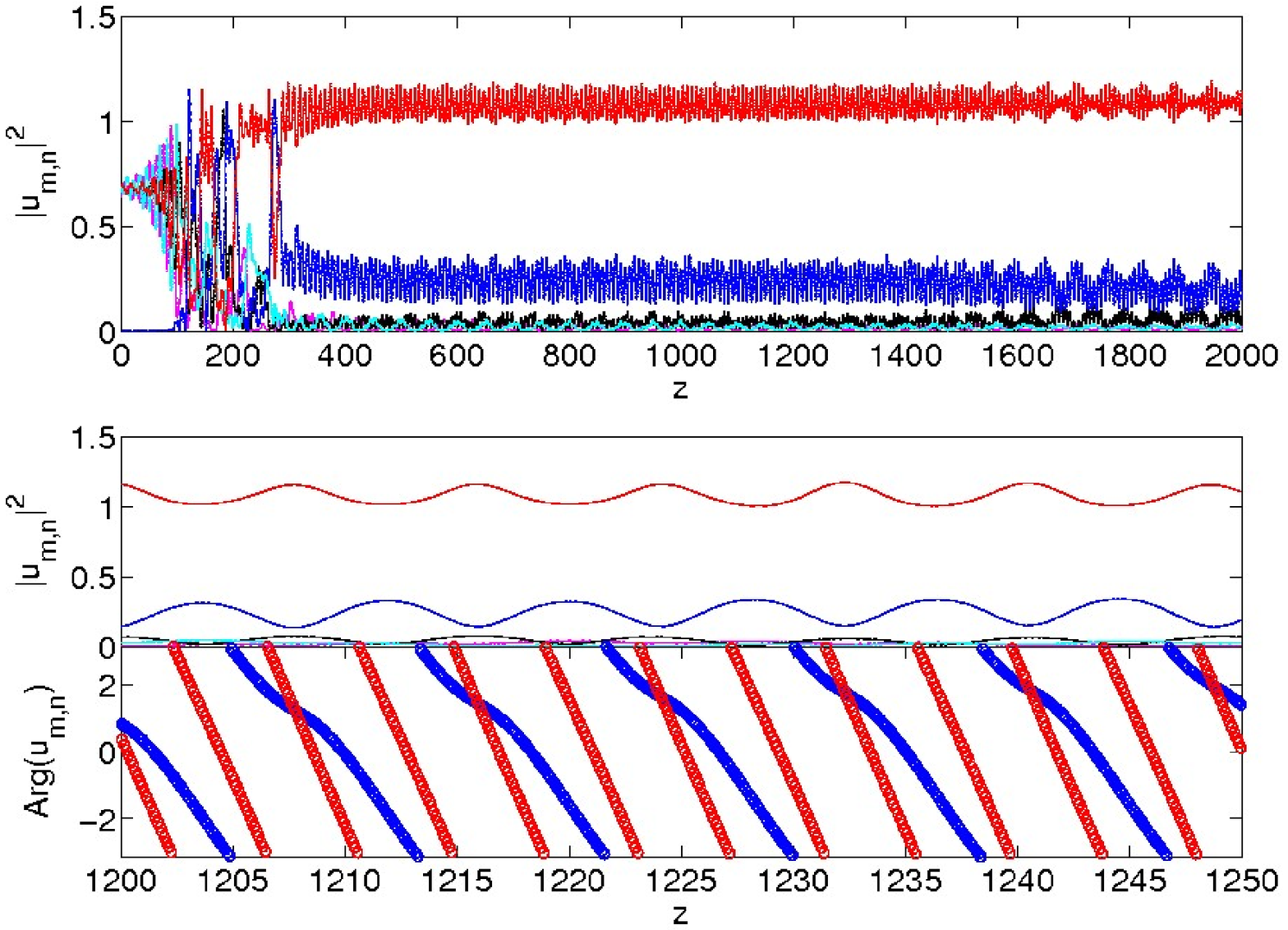}\\
\end{center}
\caption{(Color online) The same panels as Fig. \ref{ip6c},
but for the in-phase/out-of-phase quadrupole. }
\label{ipop4c}
\end{figure}

%\begin{figure}[tbp!]
%\begin{center}
%\includegraphics[width=.55\textwidth]{fig21.eps.jpg.ps}
% \end{center}
%\caption{(Color online) The same as Fig. \ref{ip6dy} but for the
%in-phase/out-of-phase quadrupole from Fig. \ref{ipop4c} (b).}
%\label{ipop4dy}
%\end{figure}

\subsection{Quadrupoles in the hourglass cell}

Next, we consider configurations around the outer 4-site contour
of the hourglass cell. Now, in this case, the five sites
comprising the hourglass are not a simple curve, i.e. the curve
crosses itself, and since some nodes are nearest-neighbors, while
others are next-nearest, the analysis must be taken to second
order for accurate predictions of all the bifurcating eigenvalues.
%(which must be treated differently in
%the defocusing regime)
% We employ the staggering tranformations,
%although, as in the trinagular lattice, we cannot use the 2D
%version due to the symmetry.  So,
Instead, we consider only
%a 1D contour and allow
%the center node to be considered both $\pm 1$ in the staggering
%transform.  This proves to give surprisingly accurate results,
%with the exception being the case in which the neighbors we
%neglect, the
%nodes in opposite corners,
%are expected to give rise to instability, i.e. they
%comprise next-nearest neighbors in-phase and we are expecting
%an imaginary pair at higher order.  Indeed an imaginary pair does
%emerge but with very small magnitude compared to the prediction.
nearest neighbors analytically and extend previous results about
higher-order interactions to make qualitative predictions there.

\subsection{In-phase quadrupoles}

First, we consider the in-phase quadrupole.  The prediction
of the discrete model
%around the contour,
%if we ignore interactions of opposite corners,
to first order, i.e. for nearest neighbors,
is that this configuration will
%be unstable with
have double
eigenvalue pairs at $\pm 2i\sqrt{\varepsilon}$ (due to the in-phase
nearest neighbors predicted to be stable).
%, and single pairs at $0$ and $\pm \sqrt{8}\varepsilon$.
On the other hand, next-nearest neighbors which are in-phase
are expected to be unstable \cite{ourpre,ourhon}
and indeed a real pair does bifurcate as well.  The real pair
comes at higher order because it is a higher-order splitting.
%of a double
%null eigenvalue considering only the four-site contour
%and also because
%(it is a next-nearest-neighbor of the sites with which
%the instability is arising).
The results are presented in Fig.
\ref{ip4d}.  The panels are the same as for Fig. \ref{ip6d}.
%, except
%that instead of the power, the right-hand column depicts the
%real (top), and imaginary (bottom) bifurcation eigenvalues.
%The agreement is surprisingly good considering the liberty
%we took in our prediction.
The evolution in Fig. \ref{ip4di}
reveals a unique
structure with three sites very close in phase and intensity,
two nearly identical and one slightly different with its intensity
oscillating with larger amplitude opposite to the others.  One of
the sites is the originally unpopulated center site, which
becomes populated when the other two disintegrate around $z=100$.

The continuum version of this configuration disappears at the
first band edge when it collides
with a more unstable solution having the center site
populated out-of-phase to the others (see Fig. \ref{ip4c}).  The main
branch is %surprisingly
only weakly unstable
%given the results of the discrete version
from the higher order interactions
and, in
fact, the second-band solution actually becomes stable for
$\mu \lesssim 3$. In the dynamical evolution of the structure from
Fig. \ref{ip4c} (b), 
%the center site very rapidly disintegrates,
%but 
the unstable five site in-phase structure remains very robust for
a long distance with reshaped phase; see Fig. \ref{ip4dy}.

\subsection{Out-of-phase quadrupoles}

Next we consider the out-of-phase quadrupole.
%for which our
%innapropriate use of liberty
%in the theory is revealed.
The predicted eigenvalues
are the same as those of the in-phase case, multiplied by $i$.
%The real-valued ones, which are due
%to the nearest neighbors
They are fairly accurate for small $\varepsilon$
as one can see in the top
right panel of Fig. \ref{op4d}.
%, although the prediction of
%the imaginary pair fails miserably as one can see in the bottom
%right panel.  The dashed line is actually $i \varepsilon$ which is
%much smaller even than the prediction.
The dynamical evolution
of the solution given in the top row
of Fig. \ref{op4d}
%. Again there are 
reveals two pairs of uneven amplitude breathers 
with phases and amplitudes oscillating
opposite to each other as in Fig. \ref{ip6d} (not shown).
%shown in 
%Fig. \ref{op4di} reveals two pairs of breathers.
% like
%the ones in  Fig. \ref{s2di}.

This solution in the continuum version, as seen in Fig.
\ref{op4c}, is always unstable.  It collides with a structure
that has a similar phase pattern, but which is
%does not
surrounding rather than including the
original configuration.  At the point
of bifurcation the common structure they share is two rows of
opposite phase.  Again the unstable
configuration persists for a long
propagation distance, suffering merely a 
reshaping of the relative phase (see Fig.
\ref{op4dy}).

\subsection{In-phase/Out-of-phase quadrupoles}

Finally, we turn to the quadrupole solution which has its
nearest-neighbors in-phase and the next-nearest ones out-of-phase.
The theoretical prediction for the discrete model based on
the set of all possible dipole configurations always implies
stability.  Indeed
%, with our reduction and use of the staggering transform
%along the contour it
the precise first order calculation predicts
this configuration will also be stable with two pairs of
eigenvalues at $\pm i 2 \sqrt{\varepsilon}$.
%and one pair at
%$\pm i \sqrt{8}\varepsilon$.
Moreover, previous results \cite{ourpre,ourhon}
predict that next-nearest neighbors which are out-of-phase
will be stable, and all those in this configuration
(except the ones which are also nearest) are
out-of-phase.
The agreement is very good again as
given in the top right panel of Fig. \ref{ipop4d}.

The dynamical evolution of this configuration (not shown)
%Fig. \ref{ipop4di}, 
again reveals the
usual in-phase to out-of-phase uneven intensity breather pair,
as shown first in Fig. \ref{ip6di}.

The continuum version is presented in Fig. \ref{ipop4c}.  Stable first
and second band versions of the solution are identified and again there
are bifurcations at the first and second bands, and also intermediate
as well as extended solutions.  The solution that collides with the main
branch at the first band-edge (b) was propagated (not shown)
%in Fig. \ref{ipop4dy}.
%The peripheral sites break down rapidly in this case, but the original
%site along with some others 
and again the original sites persist and the relative phase %only
reshapes after $z=50$.

\section{Conclusions}
\label{conclusion}

In conclusion, we have presented results of prototypical contours
(i.e. hexagonal and ``hourglass") of localized structures in a Kagom{\'e}
lattice symmetry with a defocusing nonlinearity for both a discrete
model and an analogous continuum model that is relevant to
experiments on the nonlinear optics of photonic lattices in
photorefractive crystals.  We have identified stable configurations
such as the in-phase hexapole and single-charge six site vortex on the
honeycomb cell, as well
as the four-site in-phase/out-of-phase quadrupole and the second-gap
in-phase quadrupole on the hourglass cell.
Many of the structures admitted not only second gap localized structures
but also semi-localized structures having energy within the
second band and extended structures in the second band-gap.

The unstable solutions were evolved in time and all of the discrete structures
persisted, with recurring dynamical breathing configurations reappearing
in several cases, such as the comparable intensity in-phase pair and the
uneven  oscillating intensity pair which are out-of-phase when their
intensities are closer and in-phase when they are further.  In the continuum
model, there was very little deviation from the amplitudes of the initial
configurations and instability was manifested through phase reshaping.  This
suggests that the Kagom{\'e} lattice is a robust structure in which to perform
experiments.

This work suggests many interesting directions to pursue, the most obvious of
which is the experimental realization of these structures in optical
lattices in photorefractive media and ultracold atomic gases
\cite{something}.  Also, the structures with energy in the second band
and higher band-gaps
warrant a deeper investigation, and it is conceivable that exact breather
solutions could be found with structure similar to the ones that reappear
in the dynamics.  Beyond that, higher dimensional extensions would also be
a challenging endeavor.

\vspace{5mm}

{\it Acknowledgments}.
KJHL gratefully acknowledges the warm hospitality
of the Center for Nonlinear Studies at Los
Alamos National Laboratories.
PGK acknowledges support from NSF-DMS-0619492,
NSF-DMS-0806762, NSF-CAREER, as well as from the Alexander von
Humboldt Foundation.

\end{document}